\documentclass[aps,prd,preprint,superscriptaddress,showpacs,amsmath,amssymb]{revtex4}

\usepackage{graphicx}          
\usepackage{dcolumn}           
\usepackage{bm}                

\begin{document}


\newcommand{\PPBAR}{\ensuremath{p\bar{p}}}
\newcommand{\TTBAR}{\ensuremath{t\bar{t}}}
\newcommand{\QQBAR}{\ensuremath{q\bar{q}}}
\newcommand{\QQBARP}{\ensuremath{q\bar{q}\,^{\prime}}}
\newcommand{\WJETS}{\ensuremath{W\!(\rightarrow\!\ell\nu)\!+\!{\rm jets}}}

\newcommand{\MET}{\mbox{$\not$${E_T}$}\,}
\newcommand{\MEX}{\mbox{$\not$${E_x}$}\,}
\newcommand{\MEY}{\mbox{$\not$${E_y}$}\,}

\newcommand{\PTTZ}{\ensuremath{p^{t\bar{t}}_{z}}}

\newcommand{\GEVCC}{\ensuremath{{\rm GeV}/c^2}}
\newcommand{\GEVC}{\ensuremath{{\rm GeV}/c}}

\newcommand{\INVPB}{\ensuremath{{\rm pb^{-1}}}}

\newcommand{\DZERO}{D\O}

\newcommand{\ALPGEN}{\texttt{ALPGEN}}
\newcommand{\HERWIG}{\texttt{HERWIG}}
\newcommand{\PYTHIA}{\texttt{PYTHIA}}


\newcommand{\DILLUM}{\ensuremath{340}}             
\newcommand{\DILEVT}{\ensuremath{33}}              

\newcommand{\LTRKLUM}{\ensuremath{360}}            
\newcommand{\LTRKEVT}{\ensuremath{46}}             

\newcommand{\NWASOL}{\ensuremath{45}}              
\newcommand{\NWAMASS}{\ensuremath{170.7}}          
\newcommand{\NWASTATPOS}{\ensuremath{+6.9}}        
\newcommand{\NWASTATNEG}{\ensuremath{-6.5}}        
\newcommand{\NWASTATAVG}{\ensuremath{6.7}}         
\newcommand{\NWASYST}{\ensuremath{4.6}}            

\newcommand{\NWAMASSSTAT}                          
   {\ensuremath{\NWAMASS^{
                \NWASTATPOS}_{
                \NWASTATNEG}}}

\newcommand{\NWAMASSSYST}                          
   {\ensuremath{\NWAMASS^{
                \NWASTATPOS}_{
                \NWASTATNEG}~{\rm (stat.)} \pm
                \NWASYST~{    \rm (syst.)}}}

\newcommand{\KINSOL}{\ensuremath{30}}              
\newcommand{\KINMASS}{\ensuremath{169.5}}          
\newcommand{\KINSTATPOS}{\ensuremath{+7.7}}        
\newcommand{\KINSTATNEG}{\ensuremath{-7.2}}        
\newcommand{\KINSTATAVG}{\ensuremath{7.4}}         
\newcommand{\KINSYST}{\ensuremath{4.0}}            

\newcommand{\KINMASSSTAT}                          
   {\ensuremath{\KINMASS^{
                \KINSTATPOS}_{
                \KINSTATNEG}}}

\newcommand{\KINMASSSYST}                          
   {\ensuremath{\KINMASS^{
                \KINSTATPOS}_{
                \KINSTATNEG}~{\rm (stat.)} \pm
                \KINSYST~{    \rm (syst.)}}}

\newcommand{\PHISOL}{\ensuremath{33}}              
\newcommand{\PHIMASS}{\ensuremath{169.7}}          
\newcommand{\PHISTATPOS}{\ensuremath{+8.9}}        
\newcommand{\PHISTATNEG}{\ensuremath{-9.0}}        
\newcommand{\PHISTATAVG}{\ensuremath{9.0}}         
\newcommand{\PHISYST}{\ensuremath{4.0}}            

\newcommand{\PHIMASSSTAT}                          
   {\ensuremath{\PHIMASS^{
                \PHISTATPOS}_{
                \PHISTATNEG}}}

\newcommand{\PHIMASSSYST}                          
   {\ensuremath{\PHIMASS^{
                \PHISTATPOS}_{
                \PHISTATNEG}~{\rm (stat.)} \pm
                \PHISYST~{    \rm (syst.)}}}

\newcommand{\COMMASS}{\ensuremath{170.1}}          
\newcommand{\COMSTAT}{\ensuremath{6.0}}            
\newcommand{\COMSYST}{\ensuremath{4.1}}            

\newcommand{\COMEVENTS}{\ensuremath{24}}           
\newcommand{\COMUNION}{\ensuremath{55}}            
\newcommand{\COMOVERLAP}{\ensuremath{44\%}}        

\newcommand{\COMMASSSTAT}                          
   {\ensuremath{\COMMASS \pm
                \COMSTAT~{\rm (stat.)}}}

\newcommand{\COMMASSSYST}                          
   {\ensuremath{\COMMASS \pm
                \COMSTAT~{\rm (stat.)} \pm
                \COMSYST~{\rm (syst.)}}}


\title
{
   Measurement of the Top Quark Mass
   using Template Methods \\
   on Dilepton Events 
   in \PPBAR\ Collisions at $\bf\sqrt{s} = 1.96$ TeV
}

\affiliation{Institute of Physics, Academia Sinica, Taipei, Taiwan 11529, Republic of China} 
\affiliation{Argonne National Laboratory, Argonne, Illinois 60439} 
\affiliation{Institut de Fisica d'Altes Energies, Universitat Autonoma de Barcelona, E-08193, Bellaterra (Barcelona), Spain} 
\affiliation{Baylor University, Waco, Texas  76798} 
\affiliation{Istituto Nazionale di Fisica Nucleare, University of Bologna, I-40127 Bologna, Italy} 
\affiliation{Brandeis University, Waltham, Massachusetts 02254} 
\affiliation{University of California, Davis, Davis, California  95616} 
\affiliation{University of California, Los Angeles, Los Angeles, California  90024} 
\affiliation{University of California, San Diego, La Jolla, California  92093} 
\affiliation{University of California, Santa Barbara, Santa Barbara, California 93106} 
\affiliation{Instituto de Fisica de Cantabria, CSIC-University of Cantabria, 39005 Santander, Spain} 
\affiliation{Carnegie Mellon University, Pittsburgh, PA  15213} 
\affiliation{Enrico Fermi Institute, University of Chicago, Chicago, Illinois 60637} 
\affiliation{Joint Institute for Nuclear Research, RU-141980 Dubna, Russia} 
\affiliation{Duke University, Durham, North Carolina  27708} 
\affiliation{Fermi National Accelerator Laboratory, Batavia, Illinois 60510} 
\affiliation{University of Florida, Gainesville, Florida  32611} 
\affiliation{Laboratori Nazionali di Frascati, Istituto Nazionale di Fisica Nucleare, I-00044 Frascati, Italy} 
\affiliation{University of Geneva, CH-1211 Geneva 4, Switzerland} 
\affiliation{Glasgow University, Glasgow G12 8QQ, United Kingdom} 
\affiliation{Harvard University, Cambridge, Massachusetts 02138} 
\affiliation{Division of High Energy Physics, Department of Physics, University of Helsinki and Helsinki Institute of Physics, FIN-00014, Helsinki, Finland} 
\affiliation{University of Illinois, Urbana, Illinois 61801} 
\affiliation{The Johns Hopkins University, Baltimore, Maryland 21218} 
\affiliation{Institut f\"{u}r Experimentelle Kernphysik, Universit\"{a}t Karlsruhe, 76128 Karlsruhe, Germany} 
\affiliation{High Energy Accelerator Research Organization (KEK), Tsukuba, Ibaraki 305, Japan} 
\affiliation{Center for High Energy Physics: Kyungpook National University, Taegu 702-701; Seoul National University, Seoul 151-742; and SungKyunKwan University, Suwon 440-746; Korea} 
\affiliation{Ernest Orlando Lawrence Berkeley National Laboratory, Berkeley, California 94720} 
\affiliation{University of Liverpool, Liverpool L69 7ZE, United Kingdom} 
\affiliation{University College London, London WC1E 6BT, United Kingdom} 
\affiliation{Centro de Investigaciones Energeticas Medioambientales y Tecnologicas, E-28040 Madrid, Spain} 
\affiliation{Massachusetts Institute of Technology, Cambridge, Massachusetts  02139} 
\affiliation{Institute of Particle Physics: McGill University, Montr\'{e}al, Canada H3A~2T8; and University of Toronto, Toronto, Canada M5S~1A7} 
\affiliation{University of Michigan, Ann Arbor, Michigan 48109} 
\affiliation{Michigan State University, East Lansing, Michigan  48824} 
\affiliation{Institution for Theoretical and Experimental Physics, ITEP, Moscow 117259, Russia} 
\affiliation{University of New Mexico, Albuquerque, New Mexico 87131} 
\affiliation{Northwestern University, Evanston, Illinois  60208} 
\affiliation{The Ohio State University, Columbus, Ohio  43210} 
\affiliation{Okayama University, Okayama 700-8530, Japan} 
\affiliation{Osaka City University, Osaka 588, Japan} 
\affiliation{University of Oxford, Oxford OX1 3RH, United Kingdom} 
\affiliation{University of Padova, Istituto Nazionale di Fisica Nucleare, Sezione di Padova-Trento, I-35131 Padova, Italy} 
\affiliation{LPNHE-Universite de Paris 6/IN2P3-CNRS} 
\affiliation{University of Pennsylvania, Philadelphia, Pennsylvania 19104} 
\affiliation{Istituto Nazionale di Fisica Nucleare Pisa, Universities of Pisa, Siena and Scuola Normale Superiore, I-56127 Pisa, Italy} 
\affiliation{University of Pittsburgh, Pittsburgh, Pennsylvania 15260} 
\affiliation{Purdue University, West Lafayette, Indiana 47907} 
\affiliation{University of Rochester, Rochester, New York 14627} 
\affiliation{The Rockefeller University, New York, New York 10021} 
\affiliation{Istituto Nazionale di Fisica Nucleare, Sezione di Roma 1, University of Rome ``La Sapienza," I-00185 Roma, Italy} 
\affiliation{Rutgers University, Piscataway, New Jersey 08855} 
\affiliation{Texas A\&M University, College Station, Texas 77843} 
\affiliation{Istituto Nazionale di Fisica Nucleare, University of Trieste/\ Udine, Italy} 
\affiliation{University of Tsukuba, Tsukuba, Ibaraki 305, Japan} 
\affiliation{Tufts University, Medford, Massachusetts 02155} 
\affiliation{Waseda University, Tokyo 169, Japan} 
\affiliation{Wayne State University, Detroit, Michigan  48201} 
\affiliation{University of Wisconsin, Madison, Wisconsin 53706} 
\affiliation{Yale University, New Haven, Connecticut 06520} 
\author{A.~Abulencia}
\affiliation{University of Illinois, Urbana, Illinois 61801}
\author{D.~Acosta}
\affiliation{University of Florida, Gainesville, Florida  32611}
\author{J.~Adelman}
\affiliation{Enrico Fermi Institute, University of Chicago, Chicago, Illinois 60637}
\author{T.~Affolder}
\affiliation{University of California, Santa Barbara, Santa Barbara, California 93106}
\author{T.~Akimoto}
\affiliation{University of Tsukuba, Tsukuba, Ibaraki 305, Japan}
\author{M.G.~Albrow}
\affiliation{Fermi National Accelerator Laboratory, Batavia, Illinois 60510}
\author{D.~Ambrose}
\affiliation{Fermi National Accelerator Laboratory, Batavia, Illinois 60510}
\author{S.~Amerio}
\affiliation{University of Padova, Istituto Nazionale di Fisica Nucleare, Sezione di Padova-Trento, I-35131 Padova, Italy}
\author{D.~Amidei}
\affiliation{University of Michigan, Ann Arbor, Michigan 48109}
\author{A.~Anastassov}
\affiliation{Rutgers University, Piscataway, New Jersey 08855}
% ...
\collaboration{CDF Collaboration}
\noaffiliation

\date{\today}

\begin{abstract}

We describe a measurement of the top quark mass from events produced 
in \PPBAR\ collisions at a center-of-mass energy of 1.96 TeV, using the
Collider Detector at Fermilab.
We identify \TTBAR\ 
candidates where both $W$ bosons from the top quarks decay into leptons
($e\nu, \mu \nu$, or $\tau \nu$) from a data sample of
\LTRKLUM~\INVPB.
The top quark mass is reconstructed in each event separately by three
different methods, which draw upon simulated distributions of the
   neutrino pseudorapidity,
   \TTBAR\ longitudinal momentum,
   or neutrino azimuthal angle
in order to extract probability distributions for the top quark mass.
For each method, representative mass distributions, or templates, are 
constructed from simulated
samples of signal and background events, and parameterized to form continuous
probability density functions.
A likelihood fit incorporating these parameterized templates is then performed 
on the data sample masses in order to derive a final top quark mass.
Combining the three template methods, taking into account correlations in
their statistical and systematic uncertainties, results in a top quark mass
measurement of
   \COMMASSSYST~\GEVCC.

\end{abstract}


\maketitle


\newpage

\clearpage
\section{\label{sec:intro} Introduction }

The top quark, the weak isospin partner of the bottom quark, was first
observed by the CDF and \DZERO\ collaborations in \PPBAR\ collisions produced
at the Fermilab Tevatron~\cite{top_discovery}.
During Run I operation from 1992 to 1995, CDF acquired 109~\INVPB\ of data
at a center-of-mass energy of $1.8$~TeV, and performed the first measurements
of top quark properties.
Since the start of Run II at the Tevatron in 2001, CDF has collected
integrated luminosities several times that of Run I.
Increased top production from a higher collision energy and improved 
acceptance of the upgraded detector have further enhanced the Run II
top quark yield.
This larger sample size allows for more precise studies of the
characteristics of the top quark.

As with all quarks, the top quark mass is not predicted by theory, and 
therefore represents a free parameter in the standard model which must
be experimentally determined.
Tevatron Run I measurements yielded a top quark mass of
   $178.0 \pm 4.3$ \GEVCC~\cite{tevrun1_mass},
approximately 40 times heavier than the next heaviest quark, the bottom quark.
Such a large mass, close to the electroweak symmetry breaking scale
   $v\!=\!(\sqrt{2} G_F)^{-1/2}\!\approx\!246$~GeV,
suggests that the top quark may play a special role in this
process~\cite{top_tmsm}.
The subsequently large contribution to quark-loop corrections of electroweak
parameters from the heavy top quark provides for
powerful tests of the standard model.
In particular, a precise measurement of the top quark mass, coupled with 
that of the $W$ boson, leads to tighter constraints on the as yet unobserved
Higgs boson~\cite{top_higgs}.

At the Tevatron, in \PPBAR\ collisions with a center-of-mass energy of
   $1.96$~TeV,
top quarks are produced mainly in \TTBAR\ pairs, through \QQBAR\ annihilation
($\sim\!85\%$) and gluon-gluon fusion.
Because of its large width and correspondingly short lifetime
   ($\sim\!10^{-25}$ s),
the top quark decays before any hadronization can take place, so that its
existence as a ``free quark'' can be studied without the complication of
lower energy QCD effects.
In the framework of the standard model, each top quark decays almost 
exclusively to an on-shell $W$ boson and a bottom quark.
The $b$ quark subsequently hadronizes into a jet of particles, while the 
$W$ decays either to a \QQBARP\ or a lepton-neutrino pair.
Thus, the decays of the $W$ bosons determine the characteristics of a \TTBAR\
event and, consequently, the event selection strategy.

The ``all hadronic'' mode, where both $W$'s decay into \QQBARP\ pairs, occurs
for
   $\sim\!44\%$ 
of \TTBAR\ events, but this topology is dominated by a large QCD multijet
background.
The most precise top quark mass measurements arise from the ``lepton+jets'' 
mode
   ($\sim\!30\%$ of events),
where one $W$ decays hadronically while the other decays to either an electron
or muon plus a neutrino, whose presence can be inferred from missing transverse
energy in the detector.
A third mode occurs when
both $W$ bosons from each top quark decay into 
leptons ($e\nu, \mu \nu$, or $\tau \nu$).
Though this ``dilepton'' mode accounts for only
   $\sim\!11\%$
of \TTBAR\ events, such measurements are important in order to
reduce the overall uncertainty on the top quark mass.
Further, dilepton measurements test the 
consistency of top quark mass results obtained using other decay modes, as
the dilepton mode contains different background sources and, therefore,
represents an inherently different event sample.
Since all top quark mass measurements assume a sample composition of \TTBAR\
and standard model background events, any discrepancy among the measured top 
masses could indicate the presence of new physics processes.

This paper reports a measurement of the top quark mass with the CDF II
detector by combining three analysis methods for the dilepton channel.
Each analysis selects candidate \TTBAR\ dilepton decays using
one of two complementary event selection strategies,
which differ in lepton identification criteria and subsequent
signal-to-background ratios.
In each analysis a single, representative top quark mass for each event
is reconstructed using different kinematical assumptions in
order to constrain the \TTBAR\ dilepton decay.
The distributions of reconstructed top quark masses obtained from the data
are compared with simulated mass distributions (templates) for signal and 
background events, and likelihood fits are used to arrive at a final top 
quark mass for each analysis technique.
Accounting for correlations in statistical and systematic uncertainties,
the results of the three analyses are then combined to determine
the top quark mass in the dilepton channel using template methods.


\clearpage
\section{\label{sec:selection} Detector and Event Selection }

\begin{figure}[tbp]
   \begin{center}
   \includegraphics[width={10cm}]{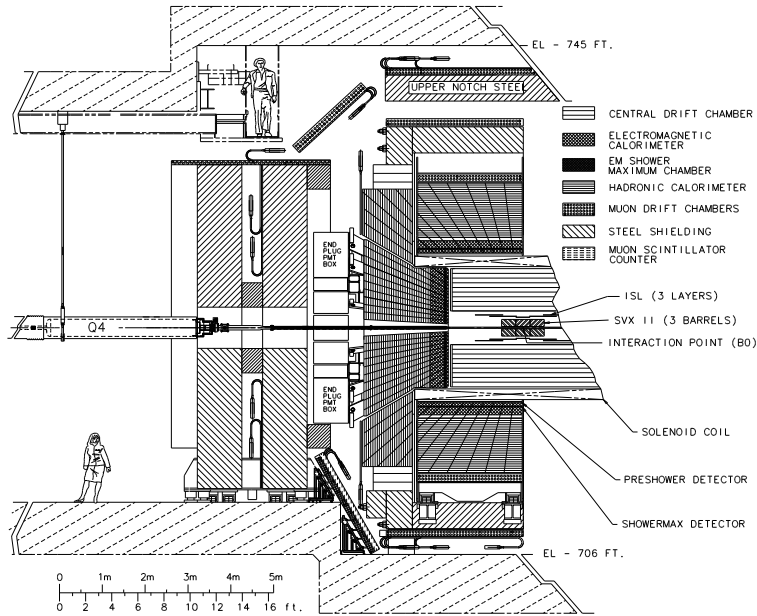}
   \caption
   {
      Elevation view of the CDF II detector, showing the inner silicon
      microstrip detector, Central Outer Tracker drift chamber, 
      electromagnetic and hadronic calorimeters, and muon drift
      chambers and scintillation counters.
   }
   \label{fig:detector}
   \end{center}
\end{figure}

The data sample used for these analyses was collected by the Collider 
Detector at Fermilab~\cite{cdfrun2_detector} during Run II operation between 
March 2002 and August 2004.
As depicted in Fig.~\ref{fig:detector},
the CDF II detector is an azimuthally and forward-backward symmetric apparatus
designed to study \PPBAR\ reactions at the Tevatron.
We use a cylindrical coordinate system about the proton beam axis in which 
$\theta$ is the polar angle, $\phi$ is the azimuthal angle, and pseudorapidity 
is defined as $\eta \equiv -\ln[\tan(\theta/2)]$.
The detector has a charged particle tracking system immersed in a 1.4 T
magnetic field, aligned coaxially with the \PPBAR\ beams.
The Run~II Silicon Vertex Detector (SVX II) and Intermediate Silicon Layer
(ISL) provide
tracking over the radial range 1.5 to 28 cm~\cite{cdfrun2_svx}.
A 3.1 m long open-cell drift chamber, the Central Outer Tracker (COT), covers
the radial range from 40 to 137 cm~\cite{cdfrun2_cot}.
The fiducial region of the silicon detector extends to pseudorapidity
   $|\eta|\!\sim\!2$,
while the COT provides coverage for
   $|\eta|\!\lesssim\!1$.

Segmented electromagnetic and hadronic sampling calorimeters surround the
tracking system and measure the energy flow of interacting particles in
the pseudorapidity range
   $|\eta|\!<\!3.6$.
This analysis uses the new end plug detectors~\cite{cdfrun2_pcal} to identify 
electron candidates with
   $1.2\!<\!|\eta|\!<\!2.0$
in addition to the central detectors~\cite{cdfrun2_ccal} for lepton candidates 
with
   $|\eta|\!<\!1.1$.
A set of drift chambers and scintillation counters~\cite{cdfrun2_muons}
located outside the central hadron calorimeters and another set behind a 
60 cm iron shield detect muon candidates with
   $|\eta|\!\lesssim\!0.6$.
Additional chambers and counters detect muons in the region
   $0.6\!\leq\!|\eta|\!\leq\!1.0$.
Gas Cherenkov counters~\cite{cdfrun2_clc} located in the 
   $3.7\!<\!|\eta|\!<\!4.7$
region measure the average number of inelastic \PPBAR\ collisions per bunch
crossing and thereby determine the beam luminosity.


The signature of \TTBAR\ decays in the dilepton channel is two jets from
the $b$ quarks, two high-momentum leptons and large missing energy (due to 
the unobserved neutrinos) from the $W$ decays, and the possibility of 
extra jets from initial or final-state radiation.
The major backgrounds for dilepton \TTBAR\ events are from Drell-Yan
dilepton production
   ($q\bar{q} \rightarrow Z/\gamma^* \rightarrow e^+ e^-, \mu^+ \mu^-,
   \tau^+ \tau^-$),
\WJETS\
events where a jet ``fakes'' the signature of the second lepton, and diboson
production
   ($WW$, $WZ$, $ZZ$).

The data are derived from inclusive lepton triggers demanding central 
electrons with transverse energy
   $E_T \equiv E \sin\theta > 18$ GeV,
or central muons with transverse momentum
   $p_T \equiv p \sin\theta > 18$~\GEVC.
Electrons in the end plug are required to have 
   $E_T > 20$ GeV.
Events must also have a missing transverse energy 
   $\MET > 15$ GeV,
calculated from the vector sum 
   $-\sum_i{E_T^i \vec{n_i}}$,
where $\vec{n_i}$ is the unit vector in the azimuthal plane which points
from the beam line to the $i^{th}$ calorimeter tower.

The top quark mass analyses described here employ one of two sets of
selection criteria developed for the \TTBAR\ cross section 
measurement in the dilepton decay channel~\cite{dil_xsec_PRL}.
The first method, referred to as the ``dilepton'' (DIL) analysis, is similar
to that used in the CDF Run I measurement~\cite{cdfrun1_dilxsec},
and requires both candidate 
leptons to be specifically identified as either electrons or muons.
A second ``lepton+track'' (LTRK) method increases the efficiency of 
the event selection (at the cost of a larger background) by requiring one
well-identified lepton (electron or muon) in conjunction with an isolated
track with large transverse momentum.
This method recovers events where leptons fall in calorimeter or muon
detector gaps,
and increases the acceptance for single prong hadronic decays of the
$\tau$ lepton from $W\rightarrow\tau\nu$
(approximately 20\% of the total LTRK acceptance, compared with 12\% for 
the DIL selection).

Both selection methods demand a ``tight'' lepton in combination with a 
``loose'' lepton of opposite charge.
Requirements for the tight lepton are identical for both methods,
but differ for the loose lepton.
Tight leptons must have well-measured tracks, based on the numbers of
silicon and drift chamber hits and reconstructed vertex position,
and have
   $E_T > 20$ GeV.
Tight leptons must also satisfy the isolation requirement that the total 
calorimeter $E_T$ within a cone
   $\Delta R \equiv \sqrt{ {\Delta\eta}^2 + {\Delta\phi}^2} = 0.4$
about the lepton trajectory not exceed 10\% of the lepton's $E_T$.
Tight electrons must have lateral and longitudinal electromagnetic 
shower profiles
in the calorimeter consistent with electrons, while tight muons must point
to muon chamber hits and have a calorimeter signature compatible with
minimum-ionizing particles.
For the DIL method, loose leptons must be well-identified electrons or
muons with no isolation requirement.
Loose DIL electrons must be central, while the muon chamber hit requirements
for loose DIL muons are relaxed.
Loose leptons in the LTRK method, in contrast, are simply required to be
well-measured and isolated tracks within
   $|\eta| < 1$ 
and having
   $p_T > 20$~\GEVC.
The LTRK loose lepton isolation is determined from the $p_T$ sum of
neighboring tracks within the cone $\Delta R = 0.4$ about the lepton track,
which must not exceed 10\% that of the lepton.

At least two jets are required per event, and are derived from looking for
clusters of energy in calorimeter towers within a cone size of 
   $\Delta R = 0.4$.
This total jet $E_T$ is corrected for non-uniformities in the response of
the calorimeter as a function of $\eta$, effects from multiple \PPBAR\
collisions, and the hadronic jet energy scale of the 
calorimeter~\cite{jetcor_NIM}.
Jets are required to have
   $|\eta| < 2.5$
and
   $E_T > 15$ GeV
for the DIL analysis,
or
   $|\eta| < 2.0$
and
   $E_T > 20$ GeV
for the LTRK method.
The two highest $E_T$ jets for each event are assumed to stem from the $b$ 
quarks; this assumption is true for
   $\sim\!70\%$
of simulated \TTBAR\ events.
For application in the top quark mass measurements, these jets are further 
corrected for energy deposited from the underlying \PPBAR\ event or lost 
outside the search cone
   $\Delta R = 0.4$.
The momentum components of each $b$ quark are then calculated from the measured
jet $E_T$ and angle by assuming a $b$~quark mass of 5.0~\GEVCC.
No explicit identification of $b$ jets is used.

The final signature of a dilepton \TTBAR\ event is missing transverse
energy \MET\ in the calorimeter.
For calorimeter tower clusters associated with an identified jet, the \MET\
vector sum uses the transverse jet energy which has been corrected for 
calorimeter response and multiple \PPBAR\ collision effects.
The \MET\ for events with an identified muon is further corrected by the 
measured muon momentum.
Dilepton \TTBAR\ events must satisfy the requirement
   $\MET > 25$ GeV.
False \MET\ may arise through mismeasurement of the leptons or jets.
Therefore, both DIL and LTRK methods require a minimum angular separation
$\Delta\phi$ between lepton or jet trajectories and the \MET\ vector.
For the DIL selection, events must have
   $\Delta\phi > 20^\circ$
for all leptons and jets if $\MET < 50$~GeV.
In the LTRK method, the \MET\ vector cannot be within $5^\circ$ of either
the tight lepton direction or the axis of the loose lepton,
and jets must have $\Delta\phi > 25^\circ$ for
events with $\MET < 50$~GeV.

The dominant source of background for both selection methods is from 
Drell-Yan $( \QQBAR \rightarrow Z/\gamma^* \rightarrow ee,\mu\mu)$ events.
These events should have no real \MET, and can only satisfy the 
selection criteria if there is mismeasurement of the lepton or jet $E_T$.
Therefore, additional selection requirements are imposed for events where
the reconstructed invariant mass of the two lepton candidates lies within
   $15$~\GEVCC\
of the $Z$ boson resonance.
For these events, the DIL method requires a ``jet significance'' of $>8$,
defined as the ratio of \MET\ to the sum of jet $E_T$ projected along
the \MET\ direction.
The LTRK method increases the \MET\ requirement to 
   $\MET > 40$~GeV
for dilepton events near the $Z$ resonance.
The DIL method further suppresses background processes by requiring that the
scalar sum of jet $E_T$, lepton $p_T$, and \MET\ (denoted by $H_T$) exceed
   $200$ GeV.

Table~\ref{tab:selections} summarizes the luminosity and expected numbers
of signal and background events for the DIL and LTRK selection methods,
along with observed results from the inclusive lepton data set.
The LTRK selection comprises a 6\% greater luminosity since it is able to
accept $e\mu$ dilepton decays when muon detectors were not operational.
The acceptance and efficiency of \TTBAR\ signal events are calculated with
a full detector simulation using
\PYTHIA~\cite{PYTHIA} Monte Carlo and assuming a production cross
section of
   6.1~pb,
corresponding to a top mass of
   178~\GEVCC~\cite{Cacciari}.
The Drell-Yan, 
\WJETS\
fakes, and diboson background acceptances are estimated using
a combination of Drell-Yan and $W+$jets data, and \PYTHIA\ and 
\ALPGEN+\HERWIG~\cite{ALPGEN,HERWIG} simulation.
The total uncertainties for expected event yields include both the
statistical uncertainties of the Monte Carlo samples used, as well as
systematic uncertainties from particle identification, jet energy measurement,
and modeling of the \TTBAR\ signal and background.
Applied to the inclusive lepton data set, the DIL selection observes
\DILEVT~events, and the LTRK selection observes \LTRKEVT~events,
representing upward fluctuations for both selections from the predicted 
numbers of events at the assumed \TTBAR\ cross section.
The DIL and LTRK data samples share \COMEVENTS\ events in common, leading to
a union of \COMUNION\ events with a \COMOVERLAP\ overlap.


\begin{table}
   \caption
   {
      Luminosity, expected \TTBAR\ signal and background (with total
      uncertainties), and observed number of events for the DIL and
      LTRK selection methods.
      A \TTBAR\ cross section of 6.1~pb is assumed, corresponding
      to a top mass of 178~\GEVCC.
   }
   \begin{ruledtabular}
   \begin{tabular}{lrr}

   & {\bf DIL} & {\bf LTRK} \\ \hline

   Luminosity        & \multicolumn{1}{c}{ \DILLUM~\INVPB }
                     & \multicolumn{1}{c}{ \LTRKLUM~\INVPB }  \\ \hline

   Expected \TTBAR   &  $15.7 \pm 1.3$  &  $19.4 \pm 1.4$ \\ \hline

   Drell-Yan         &  $ 5.5 \pm 1.2$  &  $ 8.7 \pm 3.3$ \\
   \WJETS\ fakes
                     &  $ 3.5 \pm 1.4$  &  $ 4.0 \pm 1.2$ \\
   Diboson           &  $ 1.6 \pm 0.3$  &  $ 2.0 \pm 0.4$ \\

   Total background  &  $10.5 \pm 1.9$  &  $14.7 \pm 3.6$ \\ \hline

   Total expected    &  $26.2 \pm 2.3$  &  $34.1 \pm 3.9$ \\

   Observed          &  \DILEVT         &  \LTRKEVT \\

   \end{tabular}
   \end{ruledtabular}
   \label{tab:selections}
\end{table}


\clearpage
\section{\label{sec:methods} Methods for Top Mass Measurement }

Reconstruction of the top quark mass from dilepton events involves an
underconstrained system.
For lepton+jets decays, the two components of \MET\ generated by 
the single neutrino, along with other assumptions about the \TTBAR\ event 
({\it e.g.}, equal masses for the $t$ and $\bar{t}$ quarks, and invariant 
masses of the $\ell\nu$ and \QQBARP\ systems equal to the $W$ mass) are 
enough to allow a kinematically overconstrained fit.
For dilepton \TTBAR\ events, in contrast, the measured \MET\ is due to two 
neutrinos, so that the decay assumptions are insufficient to constrain the
event.

Specifically, for each \TTBAR\ event, the kinematics are fully specified 
by 24 quantities: the four-momenta of the six final state particles.
Twelve three-momentum components of the two $b$-quarks and two leptons are
measured by the detector, along with the two components of \MET.
The four mass values of the final state $b$-quarks and leptons are known,
while the two neutrinos are assumed to be massless.
Making three additional assumptions about the \TTBAR\ and $W$ boson
decays:
\begin{eqnarray}
   m(b \ell^+ \nu) &=& m(\bar{b} \ell^- \bar{\nu}) \\
   m(\ell^+ \nu) &=& m(W^+) \\
   m(\ell^- \bar{\nu}) &=& m(W^-)
\end{eqnarray}
results in only 23 measured, known, or assumed components of the system.
Therefore, the top quark mass cannot be directly reconstructed from \TTBAR\
dilepton decays, but requires one additional kinematic assumption to constrain
the system.

In practice, for each event we integrate over undetermined kinematical
variables to obtain distributions giving the relative likelihood of different 
values of the top quark mass.
The three mass analyses are distinguished by different choices of 
kinematical variable, different methods for determining the likelihood of 
each top quark mass, and different approaches to distilling the resulting 
information into one top quark mass per event.
This section describes each technique in turn.
We model the \TTBAR\ decay kinematics and optimize each method
over a large range of top quark masses,
using \HERWIG\ Monte Carlo simulation with
\texttt{CTEQ5L}~\cite{CTEQ} parton distribution functions.
Potential biases in the reconstructed top quark masses
are taken into account in the comparison of the
measured distributions with top quark mass templates
derived using the same simulation,
as discussed in Section~\ref{sec:procedure}.


\subsection{\label{sec:NWA} Neutrino Weighting Algorithm (NWA)}

One method for estimating the top quark mass from dilepton events uses
the Neutrino Weighting Algorithm (NWA).
In Run I at the Tevatron, the NWA method was one of two techniques used by
   \DZERO~\cite{d0run1_dilmass},
and was employed by
   CDF~\cite{cdfrun1_massprd}
to measure the top quark mass.
The method therefore provides a baseline for CDF Run II measurements,
and is applied to the \LTRKLUM~\INVPB\ LTRK event sample.
The strategy of the algorithm is to solve for the neutrino and antineutrino
momenta, independently of the measured missing energy, by making additional
assumptions about the \TTBAR\ decay.
The neutrino/antineutrino solutions are then compared with the measured
\MET\ through a weight function in order to create a probability
distribution for the event as a function of top quark mass.

The NWA weight function is constructed as follows.
We assume values for the top quark and $W$ boson masses, the pseudorapidities
of the neutrino and antineutrino, and the lepton-jet pairings associated with
the top/antitop decays.
We apply energy-momentum conservation to the top quark decay and obtain up to
two possible solutions for the 4-momentum
   ({\boldmath$\nu$})
of the neutrino.
We repeat this procedure on the antitop decay, resulting in up to four
possible pairs of neutrino-antineutrino solutions
   ({\boldmath$\nu,\bar{\nu}$}).
Each of the four solutions is assigned a probability (weight, $w_i$) that it
describes the observed missing transverse energy components \MEX\ and \MEY\
within their uncertainties $\sigma_x$ and $\sigma_y$, respectively:
\begin{eqnarray}
   w_i = \exp(-\frac{(\MEX - p_x^\nu - p_x^{\bar{\nu}})^2}{2\sigma_x^2}) \cdot
      \exp(-\frac{(\MEY - p_y^\nu - p_y^{\bar{\nu}})^2}{2\sigma_y^2})
      .
   \label{eqn:nwa_wi}
\end{eqnarray}
We use $\sigma_x = \sigma_y = 15$ GeV, which is obtained from $t\bar{t}$ Monte
Carlo simulation generated with $m_t = 178$ \GEVCC.
In practice, however, the performance of the algorithm is insensitive to the
particular choice of \MET\ resolution.

\begin{figure}[tbp]
   \begin{center}
   \includegraphics[width={10cm}]{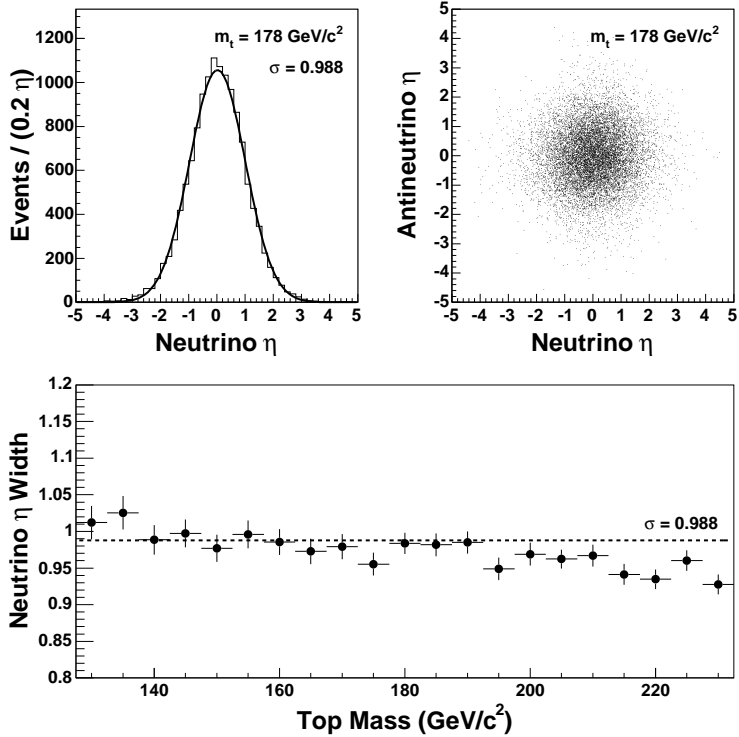}
   \caption
   {
      Neutrino $\eta$ distribution with Gaussian fit (upper left)
      and neutrino vs.\ antineutrino $\eta$ (upper right) from a
      \HERWIG\ \TTBAR\ sample with $m_t=178$ \GEVCC.
      Lower plot shows $\eta$ width as a function of generated
      top quark mass, compared with fit value at $m_t=178$ \GEVCC\
      (horizontal line).
   }
   \label{fig:nwa_neutrino_eta}
   \end{center}
\end{figure}

\begin{figure}[tbp]
   \begin{center}
   \includegraphics[width={10cm}]{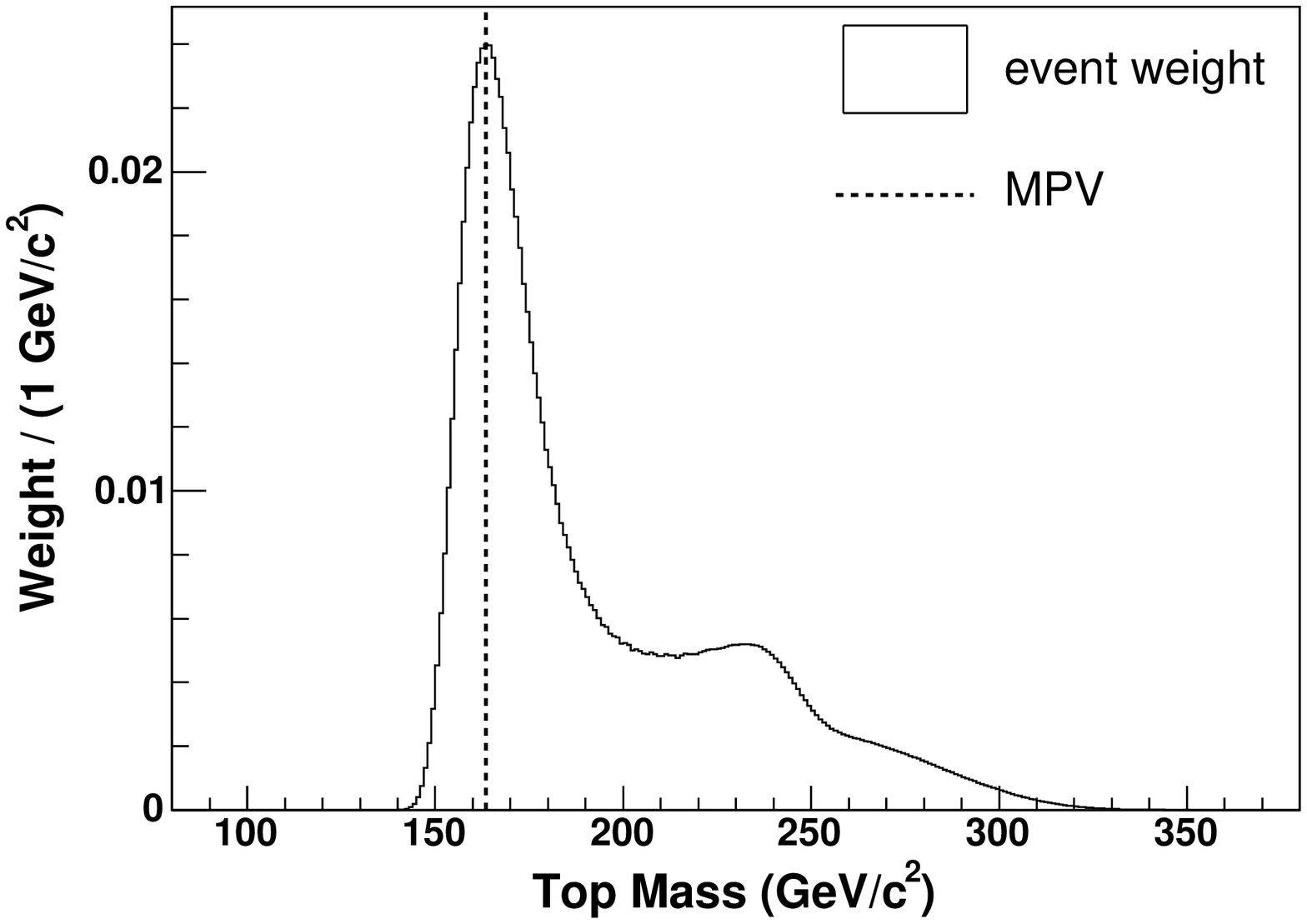}
   \caption
   {
      NWA weight distribution as a function of top quark mass hypothesis
      (from Eq.~\ref{eqn:nwa_weight}) for a \HERWIG\ Monte Carlo \TTBAR\
      event with $m_t=170$ \GEVCC.
      The vertical line denotes the most probable value (MPV) of $m_t$
      chosen by the method.
   }
   \label{fig:nwa_mc_event_weight}
   \end{center}
\end{figure}

Given the assumed top quark mass and assumed neutrino $\eta$ values, any of the 
four solution pairs ({\boldmath$\nu, \bar{\nu}$})
have {\it a priori} equal probability.
We therefore sum the four weights:
\begin{eqnarray}
   w(m_{t}, \eta_{\nu},\eta_{\bar{\nu}}, \ell\mathrm{-}jet) 
      = \sum_{i=1}^{4} w_{i}
      .
   \label{eqn:sumsolutions} 
\end{eqnarray}
Not knowing which are the true neutrino $\eta$'s in our event, we repeat the
above steps for many possible ($\eta_\nu, \eta_{\bar{\nu}}$) pairs.
As seen in the upper plots of Fig.~\ref{fig:nwa_neutrino_eta},
Monte Carlo $t\bar{t}$ simulation indicates that the 
neutrino $\eta$'s are uncorrelated, and follow a Gaussian 
distribution centered at zero with a width near one.
Since the neutrino $\eta$ width varies little with top quark mass (as shown
in the lower plot of Fig.~\ref{fig:nwa_neutrino_eta}), we assume a constant
width for all top quark masses corresponding to the value of 0.988 obtained 
from the $m_t = 178$ \GEVCC\ sample.
To ensure symmetry and smoothness,
we scan the neutrino $\eta$ distributions from $-3$ to $+3$ in steps of 0.1,
and each ($\eta_\nu, \eta_{\bar{\nu}}$) pair is assigned a probability of
occurrence $P(\eta_\nu, \eta_{\bar{\nu}})$ derived from 
a Gaussian of width 0.988.
Each trial ($\eta_\nu, \eta_{\bar{\nu}}$) pair contributes to the
event according to its weight (Eq.~\ref{eqn:sumsolutions})
and probability of occurrence, $P(\eta_\nu, \eta_{\bar{\nu}})$:
\begin{eqnarray}
   w(m_{t}, \ell\mathrm{-}jet)
      = \sum_{\eta_{\nu},\eta_{\bar{\nu}}} P(\eta_\nu, \eta_{\bar{\nu}}) \cdot
      w(m_{t}, \eta_{\nu},\eta_{\bar{\nu}}, \ell\mathrm{-}jet)
      .
\end{eqnarray}
Since we do not distinguish $b$ jets from $\bar{b}$ jets,
both possible lepton-jet pairings are summed.
Thus, the final weight becomes a function only of the top quark mass, after
integrating over all other unknowns:
\begin{eqnarray}
   W(m_{t}) = \sum_{\ell^+\mathrm{-}jet_1}^{\ell^+\mathrm{-}jet_2}
      w(m_{t}, \ell\mathrm{-}jet)
      .
   \label{eqn:nwa_weight}
\end{eqnarray}
We scan $m_t$ from 80 to 380 \GEVCC\ in steps of 1 \GEVCC.
Figure~\ref{fig:nwa_mc_event_weight} shows the resulting normalized weight
distribution from Eq.~\ref{eqn:nwa_weight} after applying the NWA method
to a \HERWIG\ Monte Carlo \TTBAR\ event, with a simulated top quark mass of
170 \GEVCC.
We choose one indicative top quark mass for each event, selecting
the most probable value (MPV) of the weight distribution as that
which best explains the event as a \TTBAR\ dilepton decay.

For a given event, there exists a small probability that the kinematics
of the decay will fail to produce a solution for any scanned top quark mass.
This efficiency for finding a solution is thus an additional event selection
criterion.
Studies of simulated \TTBAR\ dilepton events show that this NWA efficiency 
for signal is 99.8\%, and independent of generated top quark mass.
Applying the NWA method to Monte Carlo background samples shows that the
efficiency for finding a kinematical solution varies between sources, 
ranging from 94-100\%, with an average background efficiency of 96\%.


\subsection{\label{sec:KIN} Full Kinematic Analysis (KIN)}

A second method for determining the top quark mass in the dilepton channel,
called the Full Kinematic Analysis (KIN),
is applied to the 340 pb$^{-1}$ DIL selection sample.
The KIN method resolves the underconstrained dilepton \TTBAR\ decays by
introducing an additional equation for the longitudinal momentum of the
\TTBAR\ system, \PTTZ.
With the 6-particle final state constrained, the KIN method solves the
resulting kinematic equations numerically
to determine the top quark mass for each event.

\begin{figure}[tbp]
   \begin{center}
   \includegraphics[width={10cm}]{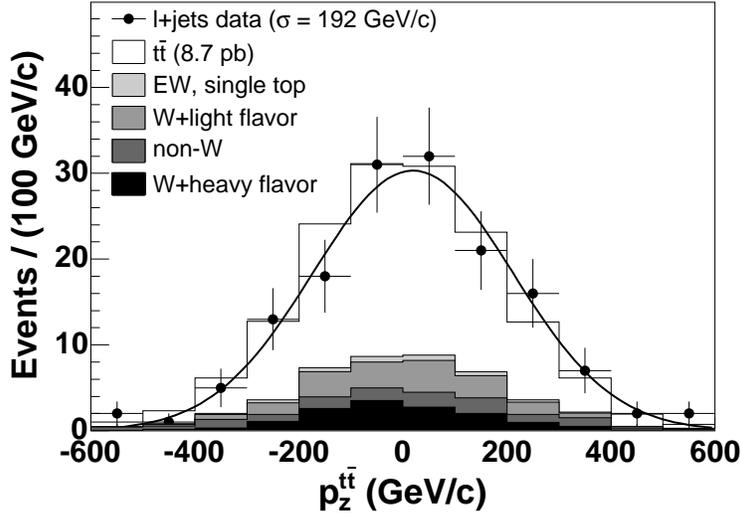}
   \caption
   {
      Comparison of \TTBAR\ longitudinal momentum between lepton+jets data
      (with associated Gaussian fit) and 
      standard model processes.
      The data corresponds to a sample of 318~\INVPB.
      The \TTBAR\ signal uses \PYTHIA\ Monte Carlo with with $m_t=178$ \GEVCC,
      corresponding to a cross section of 8.7 pb.
   }
   \label{fig:kin_pz_data}
   \end{center}
\end{figure}

Ideally, the quantity \PTTZ\ should be determined theoretically, and should 
be virtually independent of the top quark mass.
Studies from Monte Carlo simulation over a range of top quark masses from
140-200~\GEVCC\ show that \PTTZ\ has a Gaussian behavior,
with a mean of zero and a width near 180~\GEVC.
This width increases by roughly 10\% across the top quark mass range studied.
The validity of our Monte Carlo simulation can be tested with data from 
lepton+jets \TTBAR\ events, where \PTTZ\ can be reconstructed explicitly.
Figure~\ref{fig:kin_pz_data} compares \PTTZ\ from the lepton+jets data 
sample with \TTBAR\ and background Monte Carlo samples, showing good 
agreement between data and simulation.
The lepton+jets event selection, using secondary vertex $b$-quark 
identification, and subsequent backgrounds are similar to those of
the lepton+jets cross section measurement~\cite{ljets_xsec_secvtx}.

Using the measured momenta of the $b$-quarks and leptons, the two components
of the measured \MET, and assumptions about the six final-state particle 
masses, the additional constraint on \PTTZ, along with constraints on the
$W$ and \TTBAR\ decays, lead to the following set of kinematic equations:
\begin{eqnarray}
   p^{\nu}_{x} + p^{\bar{\nu}}_{x} &=& \MEX
   \label{eqn:kin_ttbar_system} \\
   p^{\nu}_{y} + p^{\bar{\nu}}_{y} &=& \MEY \nonumber \\
   p^{t}_{z} + p^{\bar{t}}_{z} &=& 0 \pm 180~\GEVC \nonumber \\
   m_{t} &=& m_{\bar{t}} \nonumber \\
   m_{W^\pm} &=& 80.4~\GEVCC \nonumber \\
   \vec{p}_{b} + \vec{p}_{W^{+}} &=& \vec{p}_{t} \nonumber \\
   \vec{p}_{\bar{b}} + \vec{p}_{W^{-}} &=& \vec{p}_{\bar{t}} \nonumber \\
   \vec{p}_{l^{+}} + \vec{p}_{\nu} &=& \vec{p}_{W^{+}} \nonumber \\
   \vec{p}_{l^{-}} + \vec{p}_{\bar{\nu}} &=& \vec{p}_{W^{-}}. \nonumber
\end{eqnarray}
These equations have two solutions, which are determined through an 
iterative procedure.
If solutions cannot be found by using the above assumptions for the
top and bottom quark masses, these requirements are relaxed, and we
accept solutions where
   $m_{W^\pm} = 80.4 \pm 3.0$~\GEVCC\ and $m_t=m_{\bar{t}} \pm 2.0$~\GEVCC.
If no solutions are found after relaxing the mass requirements,
the event is rejected.

The iterative procedure employed, Newton's Method~\cite{newton}, solves
equations of the form $f(x)$=0.
The method requires an initial guess for $x$ which is reasonably close to
the true root.
The local derivative $f'(x)$ is then computed and extrapolated to zero,
providing a better approximation for the root.
This procedure is repeated according to:
\begin{eqnarray}
   x^{n+1} = x^n-\frac{f(x^n)}{f'(x^n)}
\end{eqnarray}
until a satisfactory solution is found.
The method is extended to a system of $k$ equations
   $F(\vec{x}) = f_i(\vec{x})$
by determining the $k\!\times\!k$ Jacobian matrix
   $J_F^{ij}(\vec{x}) = \frac{\partial f_i(\vec{x})}{\partial x_j}$,
where $(i=1,k;j=1,k)$.
In actuality, the method solves the linear equations:
\begin{eqnarray}
  J_F(\vec{x}^n) \cdot (\vec{x}^{n+1}-\vec{x}^n) = -F(\vec{x}^n)
\end{eqnarray}
for the unknown $\vec{x}^{n+1}-\vec{x}^n$, in order to avoid having to
compute the inverse of $J_F(\vec{x}^n)$.

Applying Newton's Method to the \TTBAR\ decay system of 
Eq.~\ref{eqn:kin_ttbar_system},
we determine the first of two pairs of quadratic solutions for the neutrino 
momentum according to the following set of three equations:
\begin{eqnarray}
   f_{1}(p^{\nu_1}_{x},p^{\nu_1}_{y},p^{\nu_1}_{z}) &\equiv& (E_{l_{1}} + 
      E_{\nu_{1}})^2 - (\vec{p}_{l_{1}}+\vec{p}_{\nu_{1}})^2 - m_{W}^2 = 0 \\
   f_{2}(p^{\nu_1}_{x},p^{\nu_1}_{y},p^{\nu_1}_{z}) &\equiv& (E_{l_{2}} + 
      E_{\nu_{2}})^2 - (\vec{p}_{l_{2}}+\vec{p}_{\nu_{2}})^2 - m_{W}^2 = 0 \\
   f_{3}(p^{\nu_1}_{x},p^{\nu_1}_{y},p^{\nu_1}_{z}) &\equiv& (E_{l_{1}} +
      E_{\nu_{1}} + E_{b_{1}})^2 - (\vec{p}_{l_{1}} + \vec{p}_{\nu_{1}} + 
      \vec{p}_{b_{1}})^2 \\
      &-& (E_{l_{2}} + E_{\nu_{2}} +E_{b_{2}})^2 +
      (\vec{p}_{l_{2}} + \vec{p}_{\nu_{2}} + \vec{p}_{b_{2}})^2 = 0 \nonumber
\end{eqnarray}
from which the full kinematic chain is reconstructed,
and the top quark mass solutions are calculated.
The second quadratic solution for neutrino momentum
   $\vec{p}^{\;\;\prime}_{\nu_{1}} \equiv \vec{p}_{\nu_{1}} + \vec{X}$
satisfies the following set of equations:
\begin{eqnarray}
   f_{1}(x_{1},x_{2},x_{3}) &\equiv& \sqrt{m_{W}^2 + (\vec{p}_{W_{1}} +
      \vec{X})^2} - E_{l_{1}} - \sqrt{(\vec{p}_{\nu_{1}} + \vec{X})^2} = 0 \\
   f_{2}(x_{1},x_{2},x_{3}) &\equiv& \sqrt{m_{W}^2 + (\vec{p}_{W_{2}} -
      \vec{X})^2} - E_{l_{2}} - \sqrt{(\vec{p}_{\nu_{2}} - \vec{X})^2} = 0 \\
   f_{3}(x_{1},x_{2},x_{3}) & \equiv & \sqrt{(\sqrt{m_{W}^2 +
      (\vec{p}_{W_{1}} + \vec{X})^2} +E_{b_{1}})^2 - 
      (\vec{p}_{t_{1}} +  \vec{X})^2} \\
      &-& \sqrt{(\sqrt{m_{W}^2 +
      (\vec{p}_{W_{2}} - \vec{X})^2} +E_{b_{2}})^2 - 
      (\vec{p}_{t_{2}} -  \vec{X})^2} = 0 \nonumber
\end{eqnarray}
from which a second pair of top quark mass solutions is found.
Since there are two possible combinations of $b$-quark jets and leptons, 
we have a total of eight possible solutions for the top quark mass.

In order to incorporate the large range of possible \PTTZ\ values about the
mean of zero (as seen in Fig.~\ref{fig:kin_pz_data}), as well as the finite
resolutions of the measured momenta and \MET, the above procedure is 
repeated 10,000 times for each possible solution.
For each repetition, the value of \PTTZ\ is drawn from a Gaussian distribution
with zero mean and width of 180~\GEVC.
The jet energies and \MET\ are similarly smeared by Gaussians according to 
their estimated resolutions,
while the relatively better resolutions on the measured jet angles and 
lepton momenta are assumed to be perfectly measured.
Kinematic reconstruction of the smeared events results in a distribution of
possible top quark masses for a given event (consistent with the measured 
kinematic characteristics of the event and the measurement uncertainties).
The most probable value (MPV) of a spline fit to this mass distribution is 
then taken as the ``raw top quark mass'' for a given solution.

\begin{figure}[tbp]
   \begin{center}
   \includegraphics[width={10cm}]{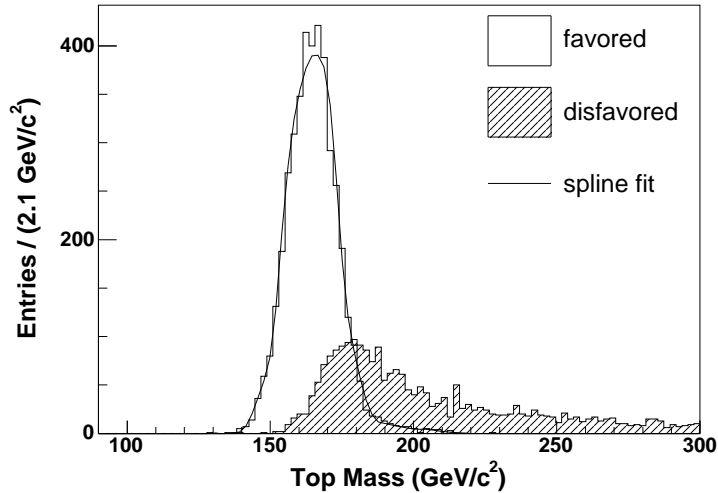}
   \caption
   {
      Smeared top quark mass distributions of the ``favored'' and
      ``disfavored'' lepton-jet pairings from the KIN method
      applied to a \HERWIG\ Monte Carlo \TTBAR\ event with $m_t=170$ \GEVCC.
      Also shown for the ``favored'' pair is the spline fit used to
      select the top quark mass for a given solution.
   }
   \label{fig:kin_mc_event_weight}
   \end{center}
\end{figure}

The KIN method then selects a single ``raw top quark mass'' from the 
eight possible solutions as follows.
Of the four possible solutions for each lepton-jet pairing, we choose that
with the smallest effective mass of the \TTBAR\ system.
Based on simulated events at $m_t = 178$ \GEVCC,
this particular mass solution is closest to the generator-level top quark
mass for approximately 84\% of the events.
The smeared mass distributions of the remaining two possible solutions (due
to the two lepton-jet pairings) are then compared, as shown in
Fig.~\ref{fig:kin_mc_event_weight} for an example simulation event.
We choose the lepton-jet pair which produces the largest number of entries
({\it i.e.} fewest number of rejections) in the smeared distribution.
The mass solution from this kinematically ``favored'' pair 
is found to be closest to the generated top quark mass for about 70\% of 
events.
In this manner, the KIN method returns a single top quark mass for each 
\TTBAR\ dilepton event.
Although this method necessarily has a bias towards
lower top masses, this bias is fully included in the 
simulation used to extract the final top mass value.

\subsection{\label{sec:PHI} Neutrino $\phi$ Weighting Method (PHI)}

A third procedure for analyzing \TTBAR\ dilepton decays, referred to as the
Neutrino $\phi$ Weighting Method (PHI),
most closely resembles the Run I lepton+jets template 
analysis~\cite{cdfrun1_massprd}.
Introducing additional assumptions about the azimuthal angle $\phi$ of the
final-state neutrinos, this method reconstructs dilepton decays through the
minimization of a chi-square functional $(\chi^2)$
to arrive at a single top quark mass for each event.
As with the KIN analysis, the PHI method uses the 340~\INVPB\
DIL selection sample.

The $\chi^2$ functional to be minimized takes the form:
\begin{eqnarray}
   \chi^2 &=&
   \sum_{\ell=1}^{2} \frac{(p_T^\ell 
       - \widetilde{p_T^\ell})^2}{{\sigma_{p_T}^\ell}^2} +
   \sum_{j=1}^{2} \frac{(p_T^j -\widetilde{p_T^j})^2}{{\sigma_{p_T}^j}^2} +
   \sum_{i=1}^{N} \frac{(UE^i -\widetilde{UE^i})^2}{{\sigma_{UE}^i}^2}
   \label{eqn:phi_chi2} \\
   &+&
   \frac{(m_{\ell_1\nu_1}-m_W)^2}{\Gamma_{W}^2} +
   \frac{(m_{\ell_2\nu_2}-m_W)^2}{\Gamma_{W}^2} +
   \frac{(m_{j_1\ell_1\nu_1}-\widetilde{m_t})^2}{\Gamma_{t}^2} +
   \frac{(m_{j_2\ell_2\nu_2}-\widetilde{m_t})^2}{\Gamma_{t}^2} \nonumber
   .
\end{eqnarray}
The first term sums over the primary lepton transverse momenta
$p_T^\ell$, with detector resolutions for the electrons and muons taken to
be~\cite{cdfrun2_detector}:
\begin{eqnarray}
   \frac{\sigma_{p_T}^e}{p_T^e}
   &=& \sqrt{\frac{0.135^2}{p_T^e}+0.02^2} \label{eqn:phi_rese} \\
   \frac{\sigma_{p_T}^\mu}{p_T^\mu}
   &=& 0.0011 \cdot p_T^\mu
   .
   \label{eqn:phi_resm}
\end{eqnarray}
The second $\chi^2$ term sums over the transverse momenta $p_T^j$ of the two
leading jets.
These transverse momenta have been further corrected for underlying event 
and out-of-cone energy, and have
a $p_T$ and $\eta$-dependent detector resolution $\sigma_{p_T}^j$
derived from simulation.
The quantity $UE$ (with uncertainty $\sigma_{UE}$) in the third $\chi^2$ 
term denotes the 
unclustered energy in the calorimeter, summed over $(i=1,N)$ towers, which is
not associated with a lepton or leading jet calorimeter cluster, but includes
any additional jets with $E_T>8$ \GEVCC\ and $|\eta |<2.5$.
The quantities $m_{\ell\nu}$ and $m_{j\ell\nu}$ in Eq.~\ref{eqn:phi_chi2} 
refer to the reconstructed invariant masses of the $W$ boson and top 
quark decay products, respectively.
For the $W$ boson decay width we use the P.D.G.\ value
   $\Gamma_{W}=2.1$~\GEVCC~\cite{PDG},
while for the top quark we assume a width of $\Gamma_{t}=2.5$~\GEVCC.
Variables with a tilde refer to the output of the minimization procedure.
The quantity $\widetilde{m_t}$ is the fit parameter returned as the
reconstructed top quark mass for the combination being considered.

To resolve the neutrino momentum used in the $W$ and top decay constraints
of Eq.~\ref{eqn:phi_chi2}, two additional assumptions are needed.
Assuming values for both neutrino azimuthal angles
   ($\phi_{\nu1},\phi_{\nu2}$),
the transverse momenta of the neutrinos are linked through the measured
\MET\ by:
\begin{eqnarray}
   p_T^{\nu1} \cdot \cos(\phi_{\nu1})+p_T^{\nu2} \cdot \cos(\phi_{\nu2})
   &=& \MEX  \\
   p_T^{\nu1} \cdot \sin(\phi_{\nu1})+p_T^{\nu2} \cdot \sin(\phi_{\nu2})
   &=& \MEY \nonumber
   \label{eqn:phi_eq1}
\end{eqnarray}
leading to the solutions:
\begin{eqnarray}
   p_x^{\nu1} \equiv p_T^{\nu1} \cdot \cos(\phi_{\nu1}) &=&
      \frac{\MEX \cdot \sin(\phi_{\nu2}) 
      - \MEY \cdot \cos(\phi_{\nu2})}{\sin(\phi_{\nu2}-\phi_{\nu1})}
      \cdot \cos(\phi_{\nu1}) \label{eqn:phi_eq2} \\
   p_y^{\nu1} \equiv p_T^{\nu1} \cdot \sin(\phi_{\nu1}) &=&
      \frac{\MEX \cdot \sin(\phi_{\nu2})
      - \MEY \cdot \cos(\phi_{\nu2})}{\sin(\phi_{\nu2}-\phi_{\nu1})}
      \cdot \sin(\phi_{\nu1}) \nonumber \\
   p_x^{\nu2} \equiv p_T^{\nu2} \cdot \cos(\phi_{\nu2}) &=&
      \frac{\MEX \cdot \sin(\phi_{\nu1})
      - \MEY \cdot \cos(\phi_{\nu1})}{\sin(\phi_{\nu1}-\phi_{\nu2})}
      \cdot \cos(\phi_{\nu2}) \nonumber \\
   p_y^{\nu2} \equiv p_T^{\nu2} \cdot \sin(\phi_{\nu2}) &=&
      \frac{\MEX \cdot \sin(\phi_{\nu1})
      - \MEY \cdot \cos(\phi_{\nu1})}{\sin(\phi_{\nu1}-\phi_{\nu2})}
      \cdot \sin(\phi_{\nu2}) \nonumber
      .
\end{eqnarray}
Performing the $\chi^2$ minimization of Eq.~\ref{eqn:phi_chi2} on all
allowed values of neutrino $\phi$ creates a set of solutions in the
$(\phi_{\nu1},\phi_{\nu2})$ plane.
In practice, only points in the quadrant
   $(0<\phi_{\nu1}<\pi, 0<\phi_{\nu2}<\pi)$
need to be sampled, since identical neutrino momentum components from
Eq.~\ref{eqn:phi_eq2} occur for the four points
   $(\phi_{\nu1}, \phi_{\nu2})$, $(\phi_{\nu1}+\pi, \phi_{\nu2})$,
   $(\phi_{\nu1}, \phi_{\nu2}+\pi)$, and $(\phi_{\nu1}+\pi, \phi_{\nu2}+\pi)$.
Since $p_T^{\nu1, \nu2}$ must be positive by definition, and will only change 
sign by adding $\pi$ to 
$\phi_{\nu1,\nu2}$, only one of the four points represents a physical
solution.
Solutions from other points are unphysical and can be interpreted as
``mirror reflections'' of the physical solution.

A grid of 12 $\times$ 12 points in the $(\phi_{\nu1},\phi_{\nu2})$ plane is
chosen, in a manner which avoids points where
   $\sin(\phi_{\nu1}-\phi_{\nu2})=0$
and Eq.~\ref{eqn:phi_eq2} becomes undefined.
At each point, 8 solutions exist due to the two-fold ambiguity in
longitudinal momentum for each neutrino, and the two possible lepton-jet
combinations.
Thus, for each event, 1152 minimizations of Eq.~\ref{eqn:phi_chi2} are
performed, each returning
an output $\chi^2$ and reconstructed top quark mass $m^{\rm rec}$.
The minimal value for $\chi^2$ among the 8 possible solutions at each point
is retained, reducing each event to an array of 144 $\chi^2_{ij}$ and 
$m^{\rm rec}_{ij}$ values,
where $(i=1,12; j=1,12)$ refer to the $(\phi_{\nu1},\phi_{\nu2})$ grid points.
Each point is weighted by its returned $\chi^2$ value according to:
\begin{eqnarray}
   w_{ij} = \frac{\exp(-\chi^2_{ij}/2)}
      {\sum_{i=1}^{12}\sum_{j=1}^{12}\exp(-\chi^2_{ij}/2)}
   \label{eqn:phi_w}
\end{eqnarray}
to create a probability density distribution normalized to unity.

\begin{figure}[tbp]
   \begin{center}
   \includegraphics[width={10cm}]{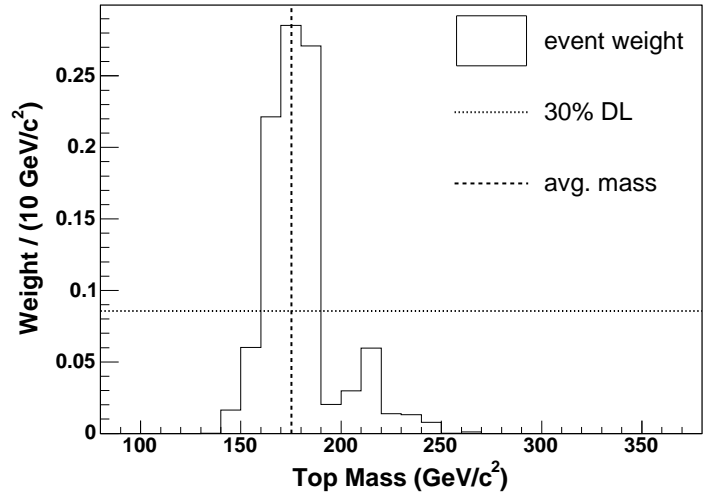}
   \caption
   {
      Binned weight distribution from the PHI method
      for a \HERWIG\ Monte Carlo \TTBAR\ event with $m_t=170$ \GEVCC,
      showing the resulting average mass for bins above the
      30\% discrimination level (DL).
   }
   \label{fig:phi_mc_event_weight}
   \end{center}
\end{figure}

To arrive at a single top quark mass value per event, the reconstructed mass 
values $m^{\rm rec}_{ij}$ of the array are averaged, using the weights derived
from Eq.~\ref{eqn:phi_w}.
The sensitivity to the top quark mass is enhanced by averaging only points with
a weight at least 30\% that of the most probable value in the probability
density distribution.
Figure~\ref{fig:phi_mc_event_weight} shows the results of the PHI method 
applied to a \HERWIG\ Monte Carlo \TTBAR\ event with $m_t=170$ \GEVCC.

\clearpage
\section{\label{sec:procedure} Template Likelihood Procedure }

The three independent measurement techniques described in 
Section~\ref{sec:methods} produce a single top quark mass for
each event in their corresponding data samples, which are mixtures of
\TTBAR\ signal and background events.
To arrive at a final top quark mass measurement, these data events are compared 
with probability density functions (p.d.f.'s) for signal and background 
within a likelihood minimization.
The p.d.f.'s are developed from template mass distributions created by applying
the NWA, KIN, and PHI methods to simulated \TTBAR\ signal and background 
samples, which are then parameterized.
For the NWA and PHI methods, this parameterization uses a combination of
Gaussian and gamma distribution terms.
Similarly, the KIN method parameterization contains a Gaussian term in
conjunction with an approximate Landau distribution.


\subsection{\label{sec:TEMP} Template construction}

\begin{figure}[tbp]
   \begin{center}
   \includegraphics[width={10cm}]{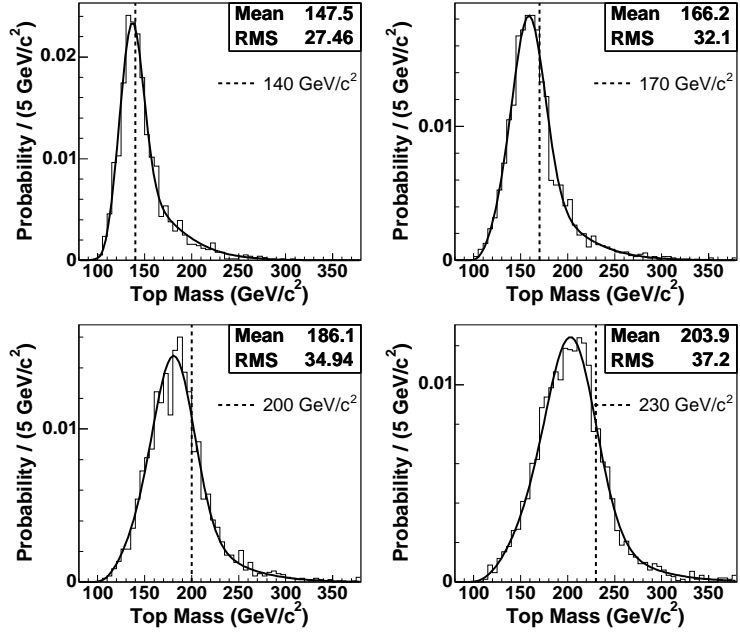}
   \caption
   {
      Example signal templates
      as a function of the reconstructed top quark mass, from
      the NWA method applied to simulated signal samples at top quark masses of
      130, 160, 190 and 220 \GEVCC.
      Overlaid are the parameterized fitting functions using
      Eq.~\ref{eqn:paramsig}.
      The vertical line indicates the generated top quark mass.
   }
   \label{fig:nwa_fittedsignal}
   \end{center}
\end{figure}

\begin{figure}[tbp]
   \begin{center}
   \includegraphics[width={10cm}]{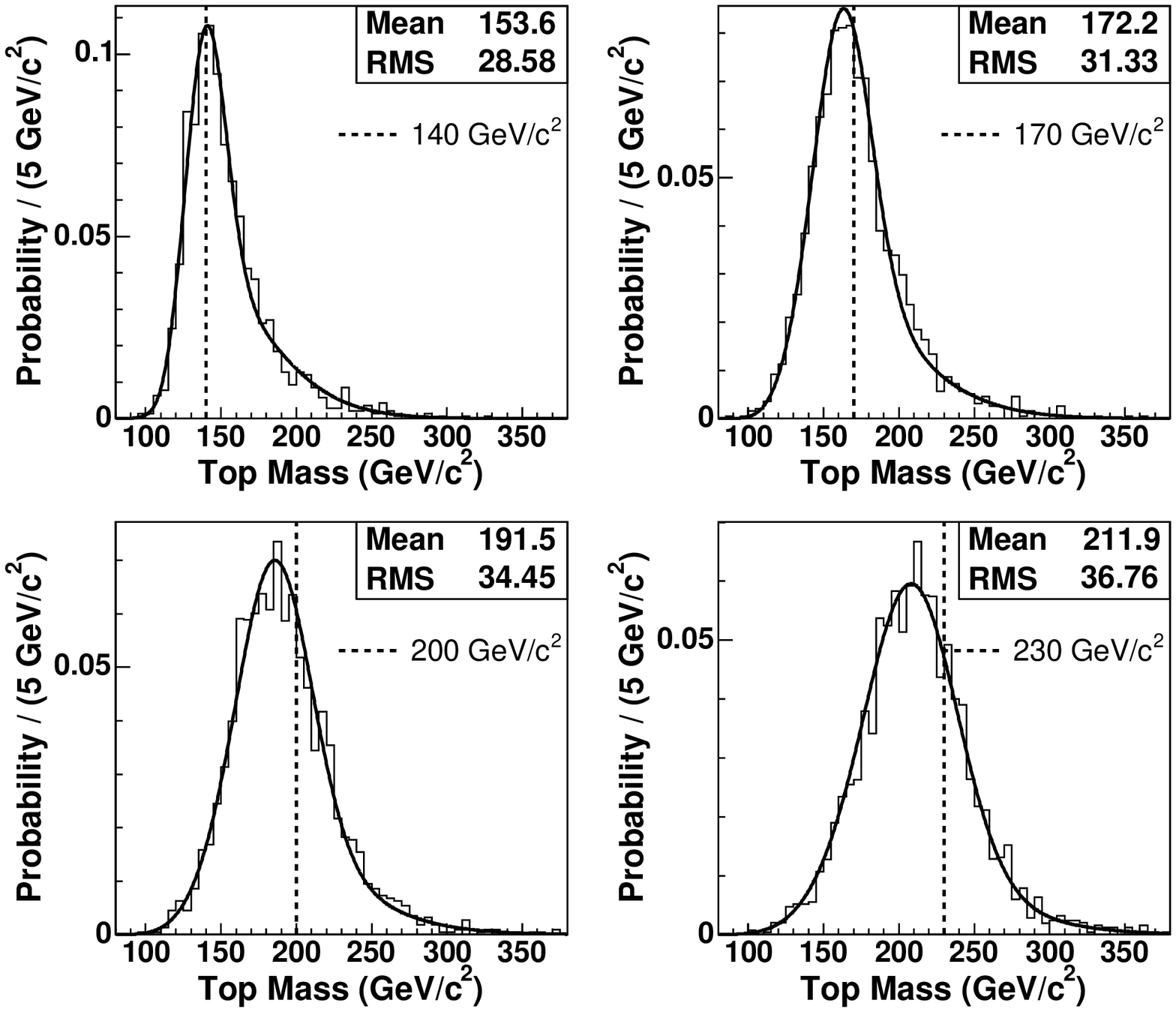}
   \caption
   {
      Example signal templates
      as a function of the reconstructed top quark mass, from
      the PHI method applied to simulated signal samples at top quark masses of
      140, 160, 190 and 220 \GEVCC.
      Overlaid are the parameterized fitting functions using
      Eq.~\ref{eqn:paramsig}.
      The vertical line indicates the generated top quark mass.
   }
   \label{fig:phi_fittedsignal}
   \end{center}
\end{figure}

For the signal, we use \TTBAR\ dilepton events generated with \HERWIG\ Monte
Carlo simulation for top quark masses from
   130 to 230 \GEVCC\ in 5 \GEVCC\
increments.
The \texttt{CTEQ5L}~\cite{CTEQ} Structure Functions are used to model the 
momentum distribution of the initial state partons.
For the NWA and PHI methods,
the signal templates obtained from this simulation are parameterized as 
the sum of a Gaussian and a gamma distribution.
This parameterization gives 
the signal p.d.f, $P_{s}(m;m_{t})$, representing the
probability of reconstructing a top quark mass $m$ when the true mass is
$m_{t}$:
\begin{eqnarray}
   P_{s}(m;m_{t})
   & = & \alpha_5 \frac{\alpha_2^{1+\alpha_1}}{\Gamma(1+\alpha_1)}
      (m-\alpha_0)^{\alpha_1} \exp{ \left( -\alpha_2(m-\alpha_0) \right) }
      \label{eqn:paramsig} \\
   & + & (1-\alpha_5)\frac{1}{\alpha_4 \sqrt{2 \pi}}
      \exp{ \left( - \frac{(m-\alpha_3)^2}{2 \alpha_4^2} \right) }
      \nonumber
   .
\end{eqnarray}
The six parameters $\alpha_i$ in Eq.~\ref{eqn:paramsig} are assumed
to be linearly dependent on the generated top quark mass, such that we in fact 
perform a 12-parameter fit for $p_i$ on all templates simultaneously, with:
\begin{eqnarray}
   \alpha_i = p_i + (m_{t}-175\,\GEVCC) \, p_{i+6}.
\end{eqnarray}
Figures~\ref{fig:nwa_fittedsignal} and~\ref{fig:phi_fittedsignal}
show representative signal templates from the NWA and PHI methods,
with the corresponding parameterized fitting function.

\begin{figure}[tbp]
   \begin{center}
   \includegraphics[width={10cm}]{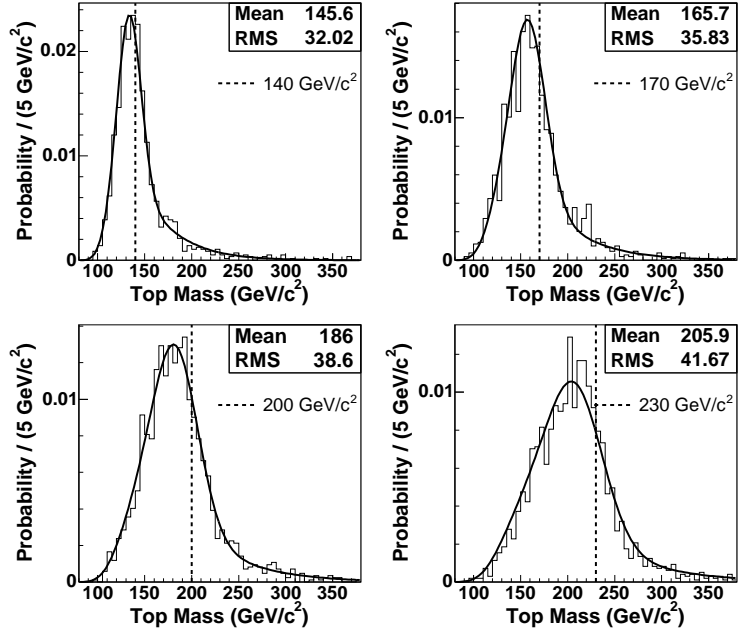}
   \caption
   {
      Example signal templates
      as a function of the reconstructed top quark mass, from
      the KIN method applied to simulated signal samples at top quark masses of
      130, 160, 190 and 220 \GEVCC.
      Overlaid are the parameterized fitting functions using
      Eq.~\ref{eqn:paramkin}.
      The vertical line indicates the generated top quark mass.
   }
   \label{fig:kin_fittedsignal}
   \end{center}
\end{figure}

The signal template parameterization employed by the KIN method contains the
Gaussian and Landau-like terms:
\begin{eqnarray}
   \lambda_{\rm Gauss}  & = &
      \frac{ m - \alpha_{4}(m_{t})}{\alpha_{5}(m_{t})}  \\
   \lambda_{\rm Landau} & = & 
      \frac{ m - \alpha_{1}(m_{t})}{\alpha_{2}(m_{t})}
\end{eqnarray}
which form the probability density function:
\begin{eqnarray}
   P_{s}(m;m_{t})
   & = &
      \frac{\alpha_{3}(m_{t})}{I_{1}} 
      \exp( -0.5 (\lambda_{\rm Landau} + \exp(-\lambda_{\rm Landau})))
      \label{eqn:paramkin} \\
   & + &
      \frac{(1-\alpha_{3}(m_{t}))}{I_{2}} \exp(-0.5 \lambda_{\rm Gauss}^{2})
      \nonumber
\end{eqnarray}
for reconstructing a top quark mass $m$ given a true mass $m_t$.
The Gaussian and Landau terms are normalized for solutions within the
reconstructed mass range $90 < m < 300$~\GEVCC\ by the integrals:
\begin{eqnarray}
   I_1 & = &
      \int_{90}^{300} e^{-\frac{1}{2}(\lambda_{\rm Landau} 
      + e^{(-\lambda_{\rm Landau})})}  \, {\mathrm{d}}m, \\
   I_2 & = &
      \int_{90}^{300} e^{-\frac{1}{2}\lambda_{\rm Gauss}^2}  \,
      {\mathrm{d}}m
   .
\end{eqnarray}
The parameters
   $\alpha_i \, (i\!=\!1,5)$
are simultaneously fit to all templates by assuming a linear dependence
on the true top quark mass $m_t$:
\begin{eqnarray}
   \alpha_i(m_{t}) = a_i + b_i * m_t
   .
\end{eqnarray}
Example signal templates using the KIN method parameterization are shown in
Fig.~\ref{fig:kin_fittedsignal}.
We observe that, for all template methods, the mean of the signal
template lies above the generated top quark mass for the 
$m_t=140$~\GEVCC\ sample, but moves below the generated value for
higher mass samples.

\begin{figure}[tbp]
   \begin{center}
   \includegraphics[width={10cm}]{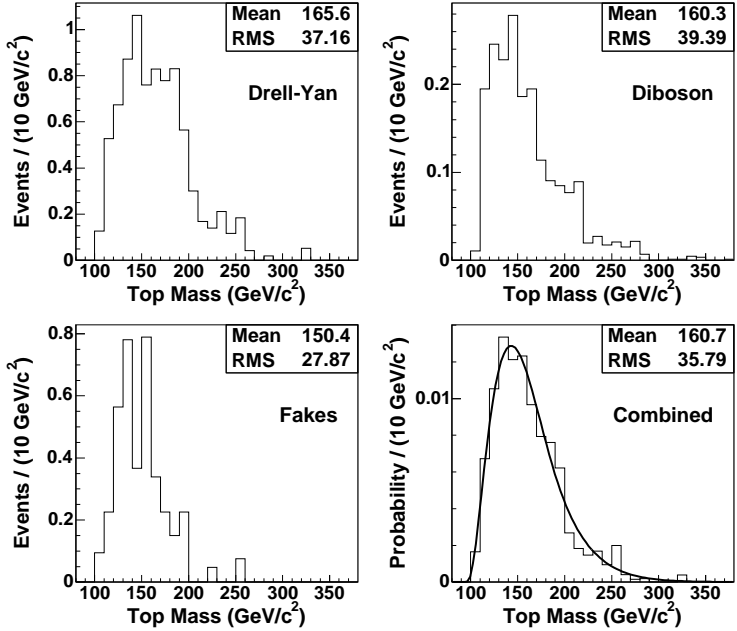}
   \caption
   {
      Reconstructed top quark mass templates for the Drell-Yan, Diboson,
      and Fakes background sources using the NWA method,
      along with the combined background template and associated 
      fitted probability density function.
      Background sources are normalized to the expected contribution in
      the \LTRKLUM~\INVPB\ LTRK sample.
   }
   \label{fig:nwa_backgrounds}
   \end{center}
\end{figure}

\begin{figure}[tbp]
   \begin{center}
   \includegraphics[width={10cm}]{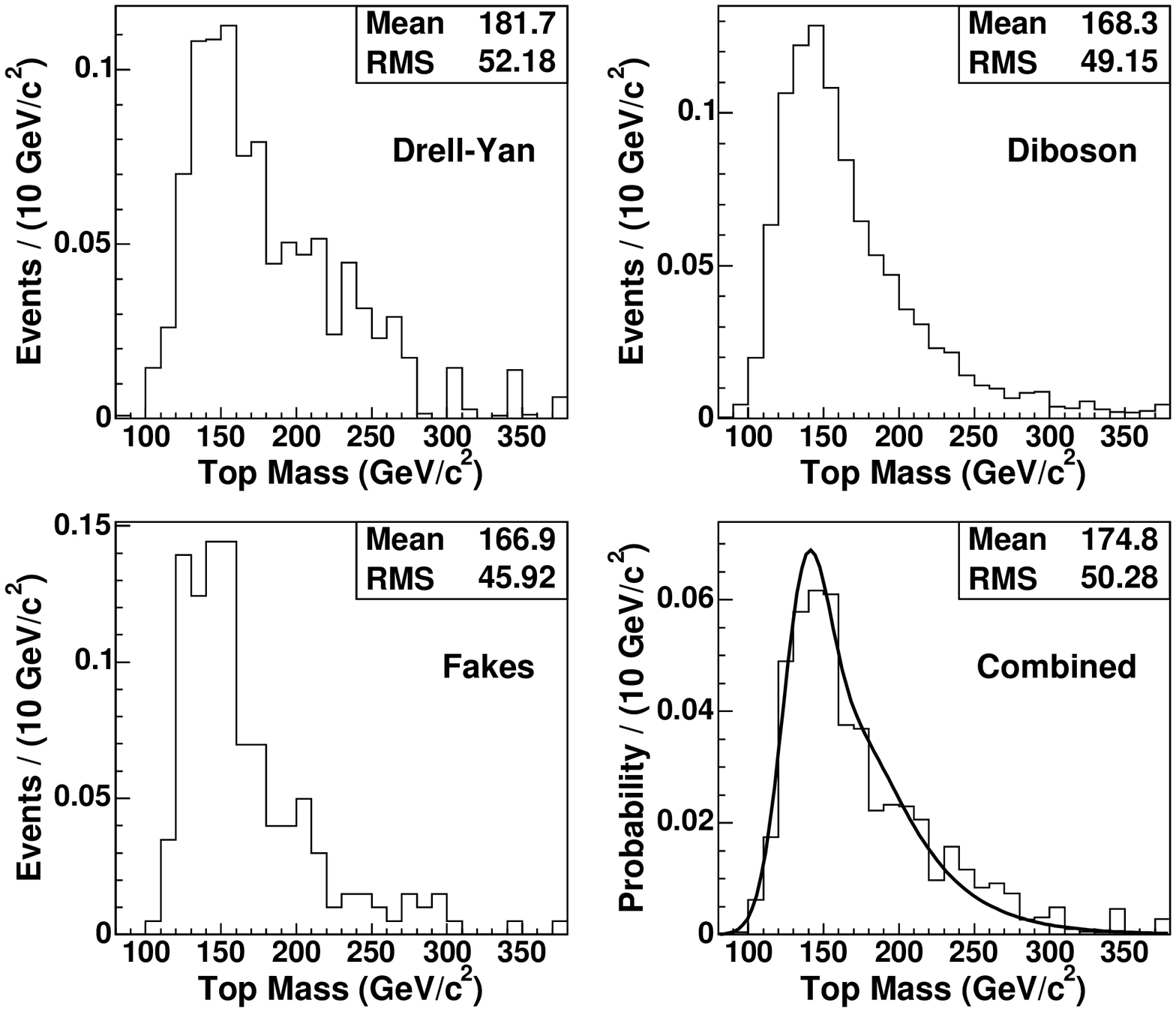}
   \caption
   {
      Reconstructed top quark mass templates for the Drell-Yan, Diboson,
      and Fakes background sources using the KIN method,
      along with the combined background template and associated
      fitted probability density function.
      Background sources are normalized to the expected contribution in
      the \DILLUM~\INVPB\ DIL sample.
   }
   \label{fig:kin_backgrounds}
   \end{center}
\end{figure}

\begin{figure}[tbp]
   \begin{center}
   \includegraphics[width={10cm}]{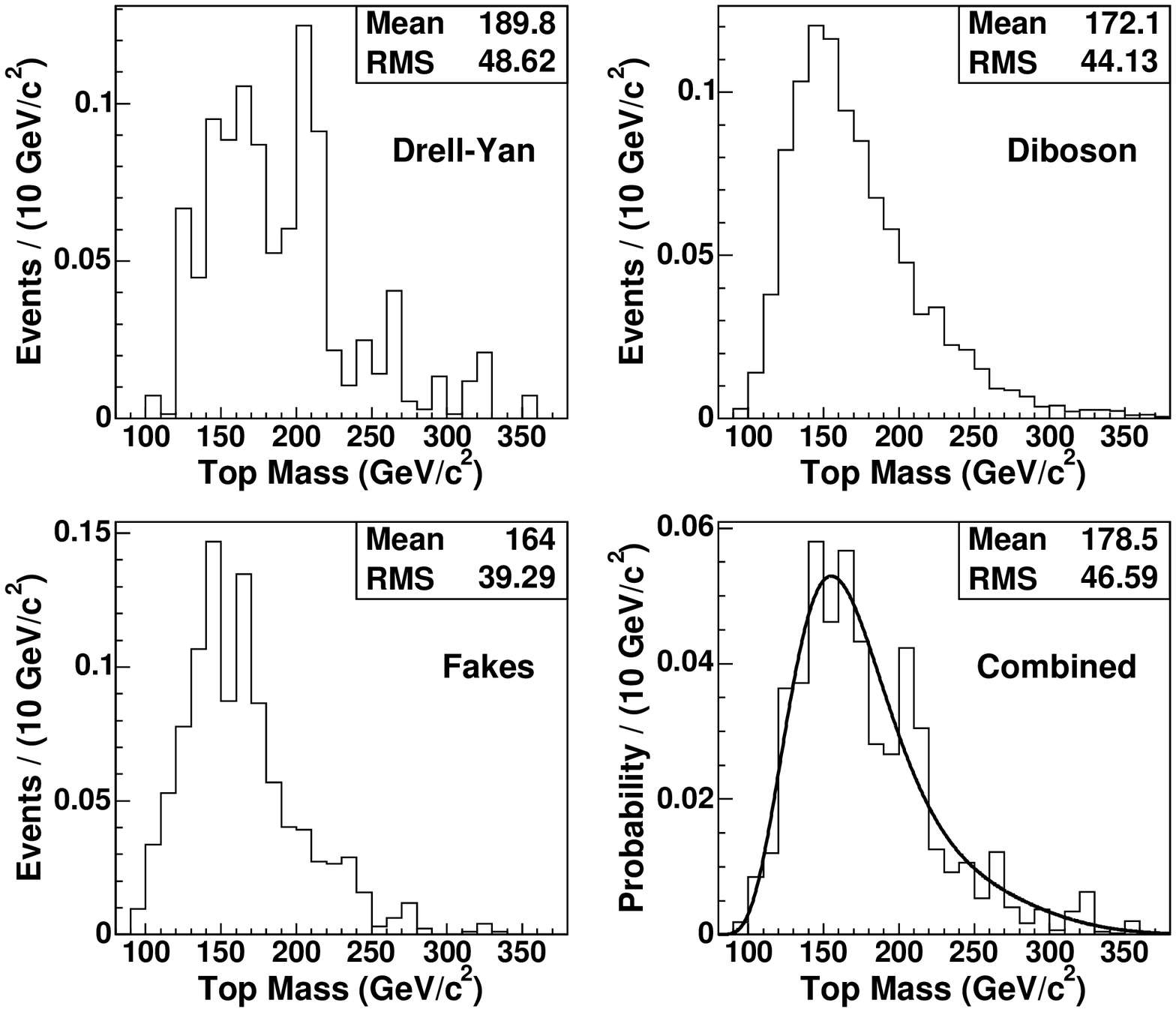}
   \caption
   {
      Reconstructed top quark mass templates for the Drell-Yan, Diboson,
      and Fakes background sources using the PHI method,
      along with the combined background template and associated
      fitted probability density function.
      Background sources are normalized to the expected contribution in
      the \DILLUM~\INVPB\ DIL sample.
   }
   \label{fig:phi_backgrounds}
   \end{center}
\end{figure}

For the background events, we create one representative background template
by adding the individual templates from each background source according
to their expected yields from Table~\ref{tab:selections}.
The templates from the various background processes are reconstructed from
fully simulated Monte Carlo samples:
   the Drell-Yan events from \PYTHIA,
   the \WJETS\ fakes from \ALPGEN+\HERWIG\ 
      simulation of $W (\to e \nu) + 3$ partons,
   and the diboson from \PYTHIA\ and \ALPGEN+\HERWIG.
In combining these sources for each mass measurement technique, the measured 
efficiencies for finding a mass solution for each simulated background source
are taken into account.
We obtain the background p.d.f.\ ($P_{b}(m)$) by fitting the combined 
background template with a functional form identical to that used for the
signal templates (Eq.~\ref{eqn:paramsig} for the NWA and PHI methods, and
Eq.~\ref{eqn:paramkin} for the KIN method),
but with fitted parameters independent of true top quark mass $m_t$.
The resulting mass templates for the three background sources,
along with the combined background template and parameterized fit,
are plotted for each method in 
Figs.~\ref{fig:nwa_backgrounds}-\ref{fig:phi_backgrounds}.

\subsection{\label{sec:LIK} Likelihood minimization}

The final step for each dilepton template analysis is the determination 
of a representative top quark mass from the data sample by performing a 
likelihood fit and minimization.
The likelihood function finds the probability that our data are described 
by an admixture of background events and dilepton \TTBAR\ decays with a 
certain top quark mass.
As input we use the top quark mass values returned by the 
particular mass measurement technique applied to the data sample, and the 
parameterized probability density functions of the signal and background 
templates derived from simulation.

The total likelihood takes the form:
\begin{eqnarray}
   {\mathcal L}(m_{t})
   = {\mathcal L}_{\rm shape}(m_{t}) \times {\mathcal L}_{n_b}
   \label{eqn:ltotal}
\end{eqnarray}
where,
\begin{eqnarray}
   {\mathcal L}_{\rm shape}(m_{t})
   = \frac{e^{-(n_{s}+n_{b})}(n_{s}+n_{b})^{N}}{N!} \prod_{i=1}^{N}
     \frac{n_{s}P_{s}(m_{i};m_{t}) +
     n_{b}P_{b}(m_{i})}{n_{s}+n_{b}}
   \label{eqn:lshape}
\end{eqnarray}
and
\begin{eqnarray}
   -\ln{\mathcal L}_{n_b} 
   = \frac{(n_{b}-n_{b}^{\rm exp})^2} {2\sigma_{n_{b}}^{2}}
   .
   \label{eqn:lbackground}
\end{eqnarray}
The likelihood returns a true top quark mass hypothesis ($m_{t}$), and
estimated numbers of signal ($n_s$) and background ($n_b$) events.
We assign a probability that each event ($i$)
looks like signal and a probability that it looks like background.
The signal and background probabilities are assigned by comparing the 
measured top quark mass values $m_i$ from the data with the parameterized 
signal and background p.d.f.'s 
   $P_s$ and $P_b$.
We find the probabilities that the 
likelihood-estimate for the number of background events $n_b$ is consistent
with our {\it a priori} estimate $n_b^{\rm exp}$, and that the 
likelihood-estimate for the total number of signal ($n_s$) and background 
events is consistent with the observed number of events $N$.
The number of background events is constrained with a Gaussian 
about $n_b^{\rm exp}$ (of width
equal to the expected background uncertainty $\sigma_{n_{b}}$), while
the sum of $n_s$ and $n_b$ is constrained with a Poisson term.
In this manner, the likelihood-estimated number of signal events is 
independent of the expected number of signal events based on an
assumed \TTBAR\ cross section.
The true top quark mass hypothesis ($m_{t}$) which minimizes
   $-\ln({\mathcal L})$
is retained.

The statistical uncertainty on $m_t$ is given by the difference between 
the minimization mass result and the mass at
   $-\ln({\mathcal L}/{\mathcal L}_{\rm max}) + 0.5$.
In the NWA and KIN analyses, uncertainty on the top quark mass from
uncertainties in the signal and background template parameterizations (due to 
limited statistics of the simulated template samples) is estimated and
included as a systematic uncertainty (see Section~\ref{sec:systematics}).
The PHI analysis incorporates this parameterization uncertainty directly 
into the top quark mass statistical uncertainty through the 
addition of a third term to the likelihood function (Eq.~\ref{eqn:ltotal}):
\begin{eqnarray}
   \mathcal{L}_{\rm param} = \exp
   \left(
      - 0.5 \{
      (\vec{\alpha}-\vec{\alpha_0})^T U^{-1} (\vec{\alpha}-\vec{\alpha_0})
    + (\vec{\beta }-\vec{\beta_0 })^T V^{-1} (\vec{\beta }-\vec{\beta_0 })
      \}
   \right)
   \label{eqn:lparam}
\end{eqnarray}
where $U$ and $V$ represent the covariance matrices of the signal and 
background parameters $\vec{\alpha}$ and $\vec{\beta}$, respectively.


\clearpage
\section{\label{sec:testing} Testing with Pseudo-experiments }

\begin{figure}[tbp]
   \begin{center}
   \includegraphics[width={10cm}]{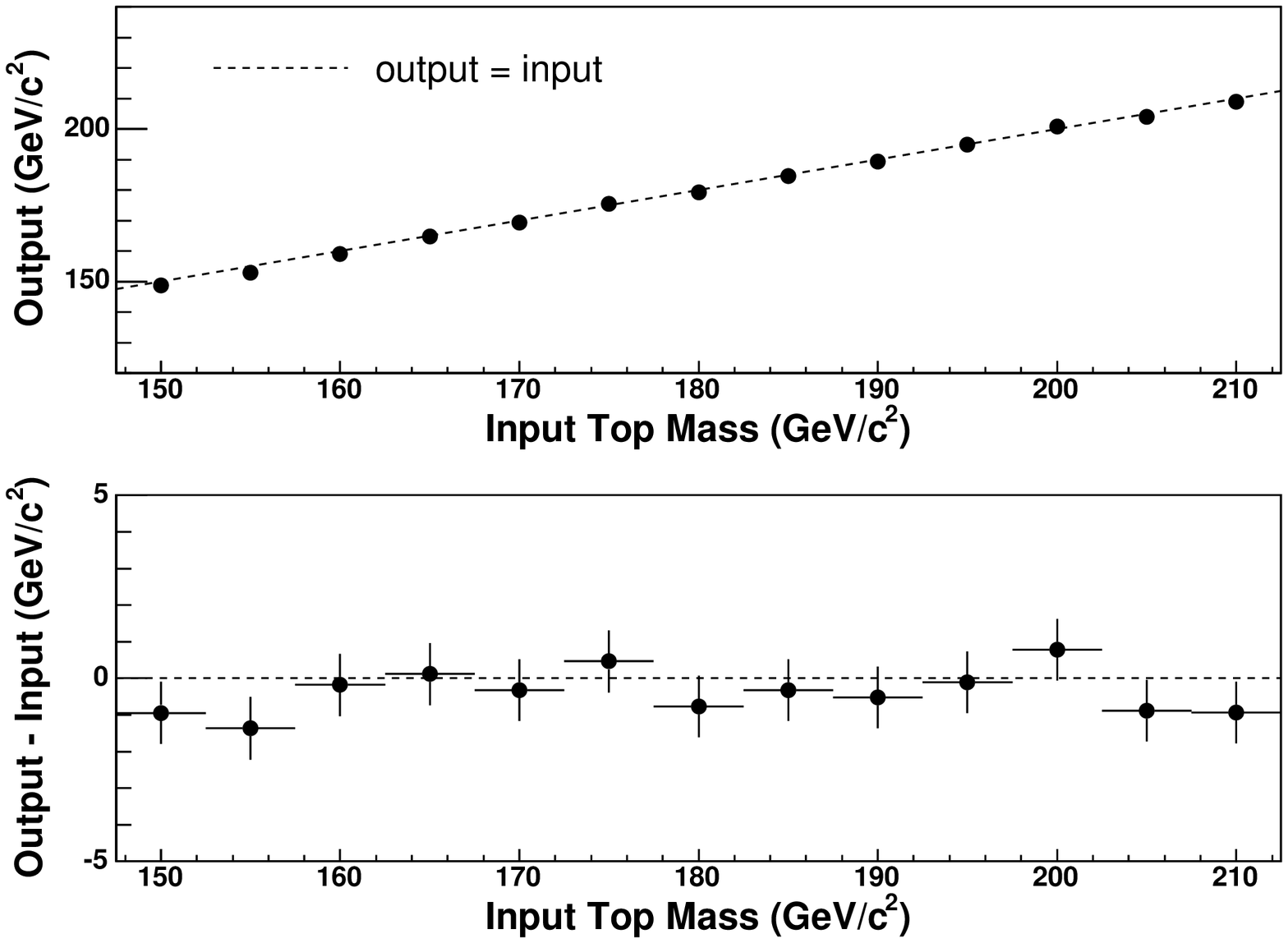}
   \caption
   {
      Results from pseudo-experiment tests of the NWA method.
      The upper plot shows the mean of the output (measured) 
      top quark mass as a function of the input (generated) mass,
      while the lower plot gives the difference between output and input
      top quark mass as a function of the input mass.
   }
   \label{fig:nwa_pe_mass}
   \end{center}
\end{figure}

\begin{figure}[tbp]
   \begin{center}
   \includegraphics[width={10cm}]{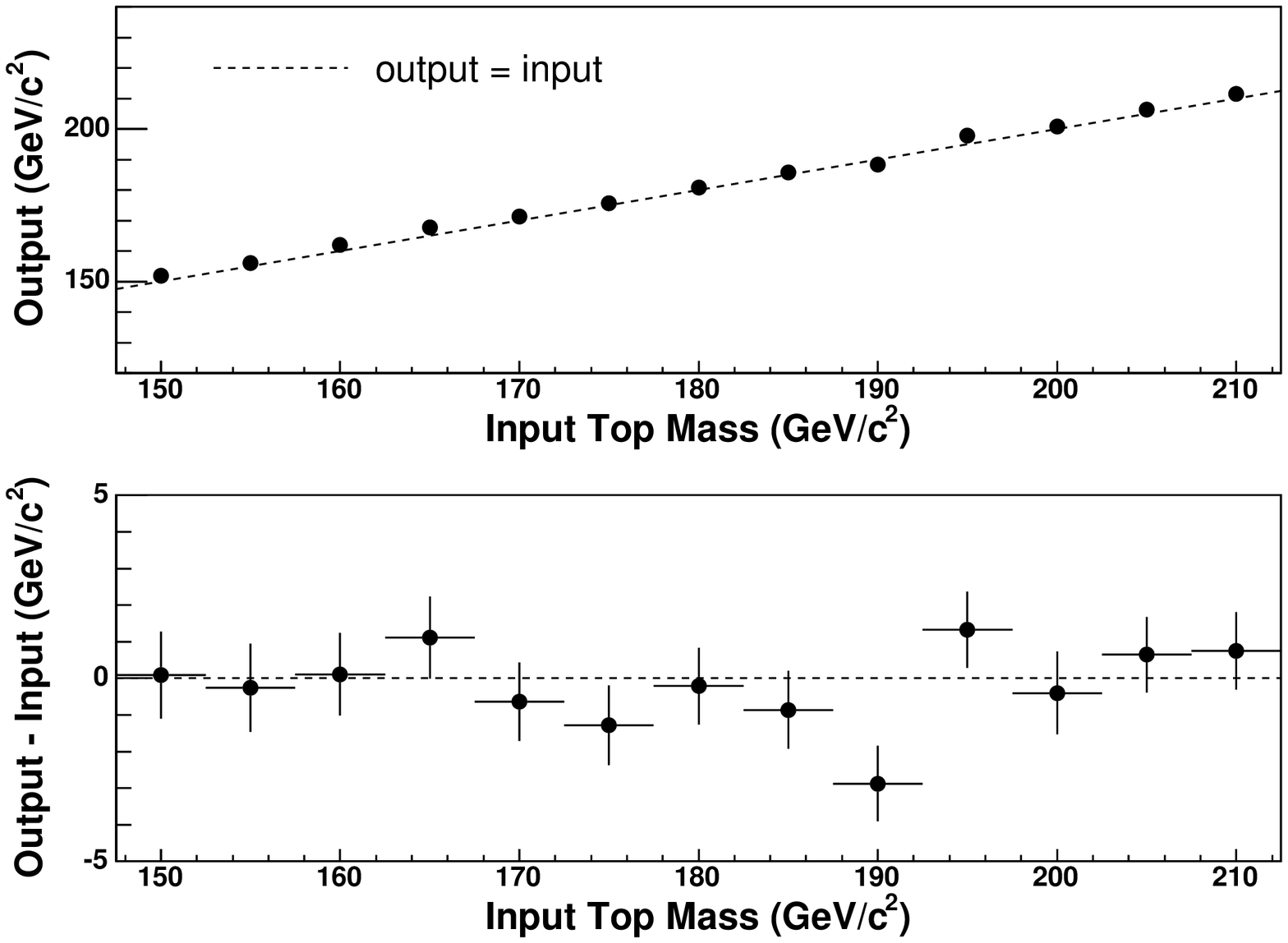}
   \caption
   {
      Results from pseudo-experiment tests of the KIN method.
      The upper plot shows the mean of the output (measured)
      top quark mass as a function of the input (generated) mass,
      while the lower plot gives the difference between output and input
      top quark mass as a function of the input mass.
   }
   \label{fig:kin_pe_mass}
   \end{center}
\end{figure}

\begin{figure}[tbp]
   \begin{center}
   \includegraphics[width={10cm}]{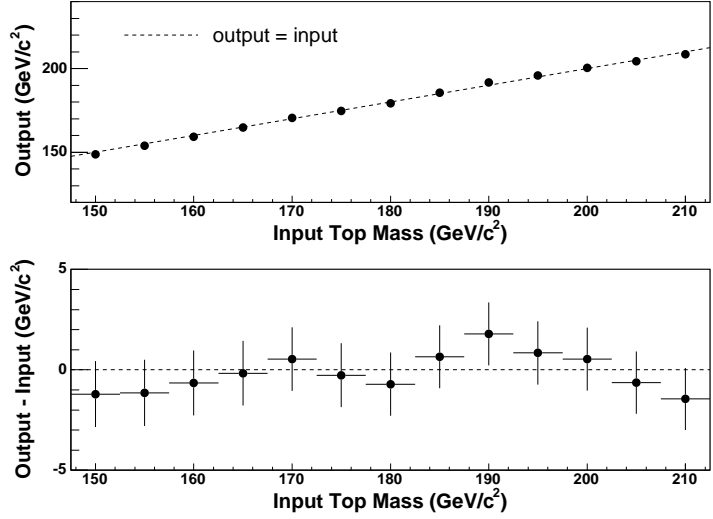}
   \caption
   {
      Results from pseudo-experiment tests of the PHI method.
      The upper plot shows the mean of the output (measured)
      top quark mass as a function of the input (generated) mass,
      while the lower plot gives the difference between output and input
      top quark mass as a function of the input mass.
   }
   \label{fig:phi_pe_mass}
   \end{center}
\end{figure}

We use a large number of simulated data ensembles, or pseudo-experiments,
to check whether the methods for mass measurement described above return the 
expected top quark mass.
For each generated top quark mass from 150 to 210 \GEVCC,
we construct a set of pseudo-experiments.
Each pseudo-experiment consists of masses drawn randomly from the signal
and background mass templates ({\it e.g.},
Figs.~\ref{fig:nwa_fittedsignal} and~\ref{fig:nwa_backgrounds}).
The numbers of signal and background events in each pseudo-experiment are
given by random Poisson fluctuations around the {\it a priori} estimates 
from the DIL and LTRK selections (see Table~\ref{tab:peevents}).
These estimates correspond to a \TTBAR\ cross section of 6.1 pb, and are
adjusted for the reconstruction efficiency of each method
for finding top quark mass solutions for signal and background events.
The likelihood minimization procedure described in the previous section
provides a ``measured'' top quark mass and statistical uncertainty for each
pseudo-experiment.
Figures~\ref{fig:nwa_pe_mass}-\ref{fig:phi_pe_mass} show the
results from these pseudo-experiments for the NWA, KIN, and PHI methods,
respectively.
The upper plots show that the measured output top quark mass tracks 
the generated input mass.
From the lower plots we observe that the residual differences between 
input and output top quark mass are consistent with zero for all 
methods, within uncertainties due to the limited statistics of the
signal and background mass templates.

\begin{table}[b]
   \caption
   {
      Expected signal and background events for the NWA, KIN, and PHI
      methods applied to the LTRK (\LTRKLUM~\INVPB) or DIL (\DILLUM~\INVPB)
      selections, and 
      corresponding to a \TTBAR\ cross section of 6.1 pb.
      Event numbers are 
      adjusted for signal and background reconstruction efficiencies
      (in parentheses).
      Also shown is the {\it a priori} statistical uncertainty on top mass
      for each method using the $m_t = 178$ \GEVCC\ simulation sample and
      correcting for underestimation found in pulls (by the scale in 
      parentheses).
   }
   \begin{ruledtabular}
   \begin{tabular}{lcrcrcrc}

   {\bf method} & {\bf luminosity}
   & \multicolumn{2}{c}{\bf expected sig.}
   & \multicolumn{2}{c}{\bf expected bkg.}
   & \multicolumn{2}{c}{\bf expected $\bm\sigma_{\bf stat}$} \\ \hline

   {\bf NWA}  &  \LTRKLUM~\INVPB
              &  $19.4 \pm 1.4$  &  (99.8\%) 
              &  $14.1 \pm 3.5$  &  (96\%)
              &  $12.8$~\GEVCC\ (1.060) \\
   {\bf KIN}  &  \DILLUM~\INVPB
              &  $12.9 \pm 1.1$  &  (75\%)
              &  $6.4 \pm 1.2$   &  (61\%)
              &  $15.1$~\GEVCC\ (1.033) \\
   {\bf PHI}  &  \DILLUM~\INVPB
              &  $17.2 \pm 1.4$  &  (100\%)
              &  $10.5 \pm 1.9$  &  (100\%)
              &  $14.5$~\GEVCC\ (1.055) \\

   \end{tabular}
   \end{ruledtabular}
   \label{tab:peevents}
\end{table}

\begin{figure}[tbp]
   \begin{center}
   \includegraphics[width={10cm}]{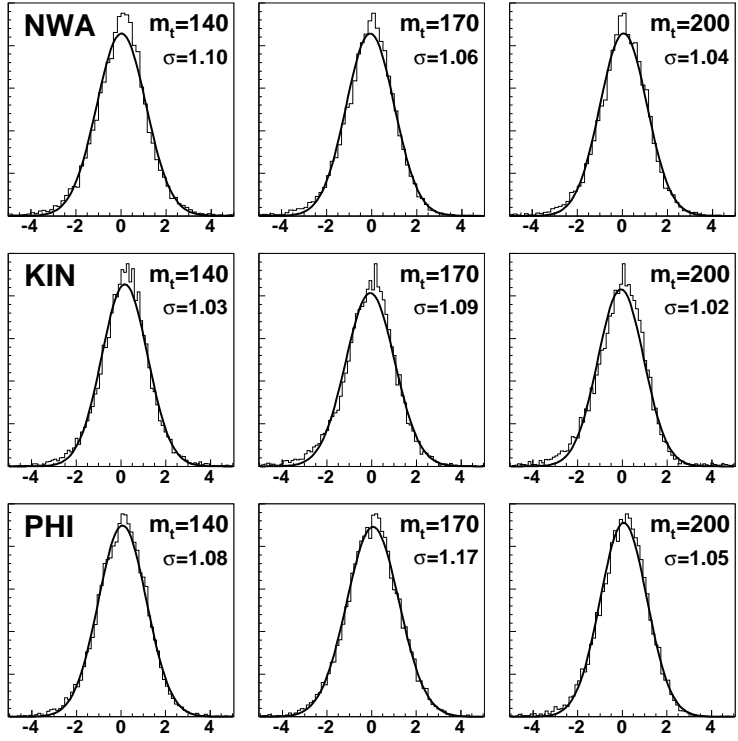}
   \caption
   {
      Example pull distributions for the NWA (upper), KIN (middle), and
      PHI (lower) pseudo-experiments,
      corresponding to generated mass samples at
         140, 170, and 200 \GEVCC.
      Each pull distribution is fit to a Gaussian (solid line), with
      noted standard deviation $(\sigma)$.
   }
   \label{fig:nwa_pull_examples}
   \end{center}
\end{figure}

\begin{figure}[tbp]
   \begin{center}
   \includegraphics[width={10cm}]{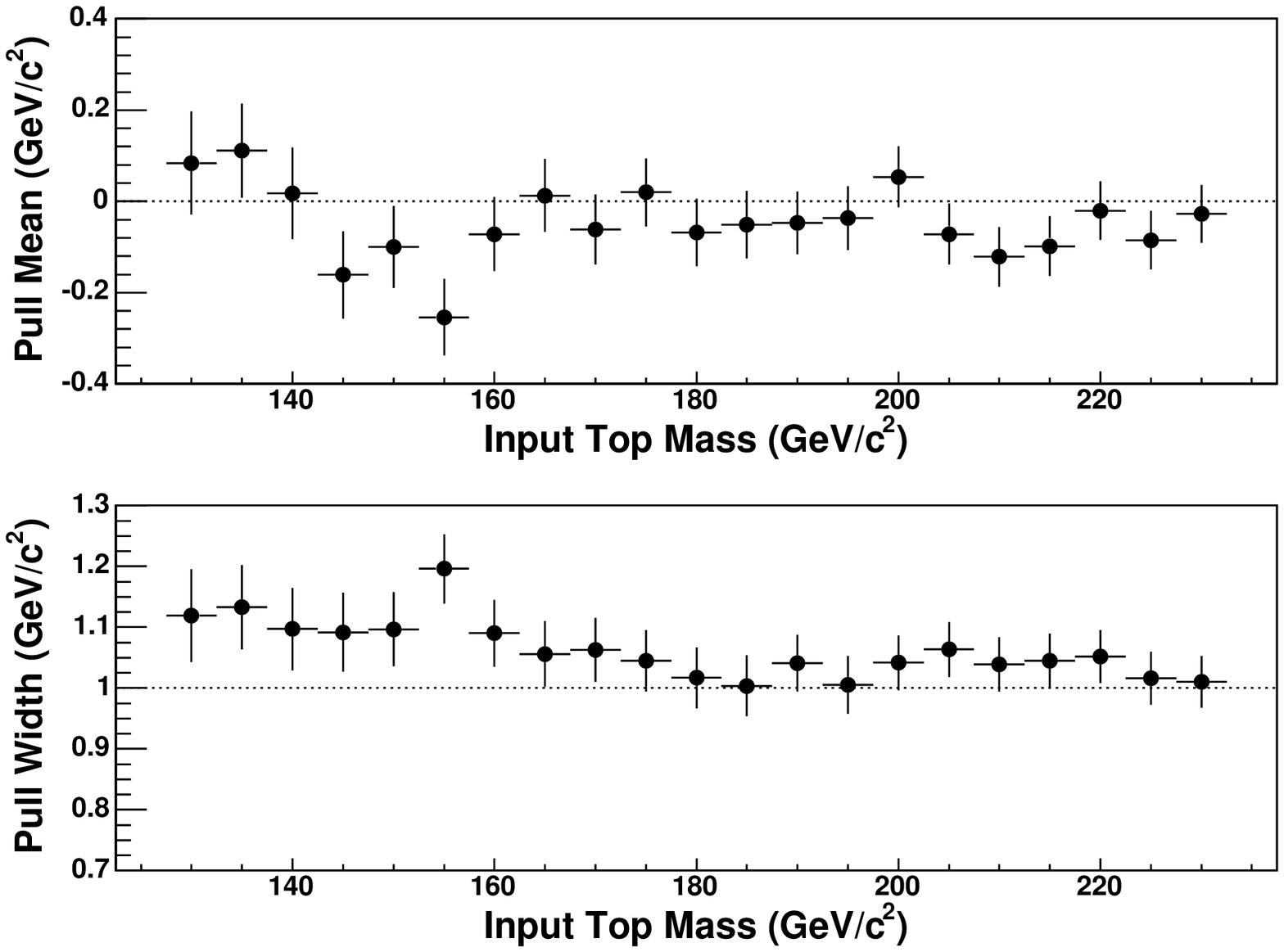}
   \caption
   {
      Summary of pull distributions for the NWA pseudo-experiments,
      showing the pull mean (upper) and width (lower) as a function
      of generated input top quark mass, compared with zero mean and
      unity width (horizontal lines).
   }
   \label{fig:nwa_pulls}
   \end{center}
\end{figure}

\begin{figure}[tbp]
   \begin{center}
   \includegraphics[width={10cm}]{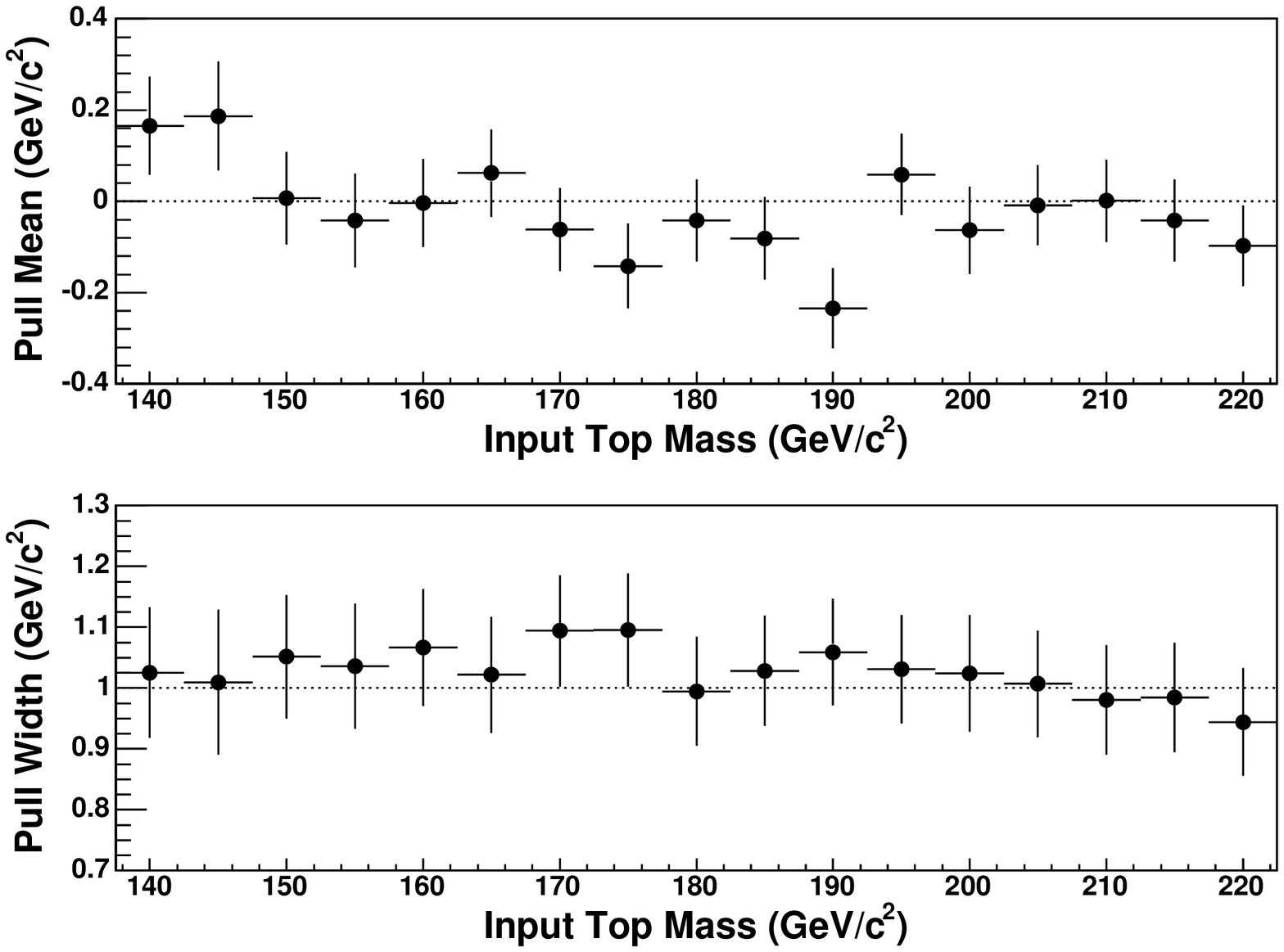}
   \caption
   {
      Summary of pull distributions for the KIN pseudo-experiments,
      showing the pull mean (upper) and width (lower) as a function
      of generated input top quark mass, compared with zero mean and
      unity width (horizontal lines).
   }
   \label{fig:kin_pulls}
   \end{center}
\end{figure}

\begin{figure}[tbp]
   \begin{center}
   \includegraphics[width={10cm}]{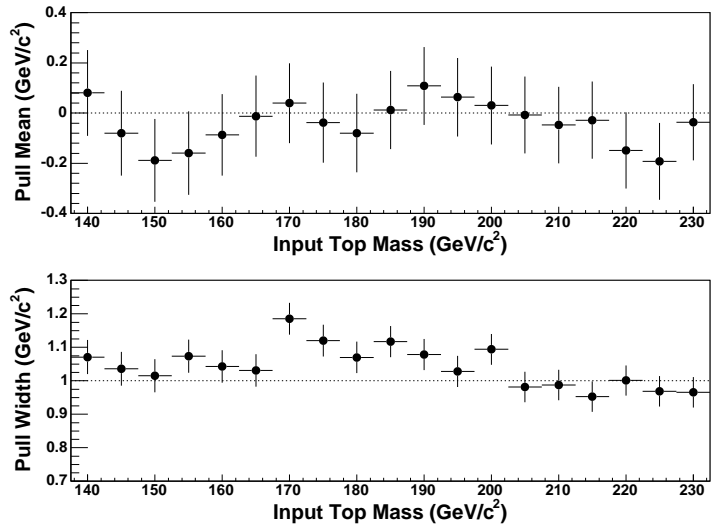}
   \caption
   {
      Summary of pull distributions for the PHI pseudo-experiments,
      showing the pull mean (upper) and width (lower) as a function
      of generated input top quark mass, compared with zero mean and
      unity width (horizontal lines).
   }
   \label{fig:phi_pulls}
   \end{center}
\end{figure}

In order to check the consistency between the spread in output
top quark mass and the estimated positive ($\sigma_{+}$) and negative
($\sigma_{-}$) statistical uncertainties from the pseudo-experiments,
pull distributions are generated according to:
\begin{eqnarray}
   pull \equiv
   \frac{m_{\rm out} - m_{\rm in}}{(\sigma_{+} \, + \, \sigma_{-})/2}
\end{eqnarray}
for each of the generated samples at different input mass (with examples shown
in Fig.~\ref{fig:nwa_pull_examples}).
Figures~\ref{fig:nwa_pulls}-\ref{fig:phi_pulls} summarize the pull mean
and width for the NWA, KIN, and PHI methods as a function of generated
top quark mass,
with corresponding uncertainties due to mass template statistics.
Non-unity widths of the pull distributions indicate that
the statistical uncertainty is underestimated for the three analyses.
Therefore, 
we scale the uncertainties obtained from the
likelihood fit on the data by the underestimation determined from the 
pseudo-experiments.
Using the $m_t = 178$~\GEVCC\ \HERWIG\ simulation and assuming
a \TTBAR\ cross section of 6.1 pb,
Table~\ref{tab:peevents} compares the expected statistical uncertainty of 
the three measurement techniques 
after applying this correction due to observed pull width (shown in 
parentheses).

\clearpage
\section{\label{sec:results} Results }

The NWA procedure is applied to the \LTRKEVT\ events satisfying the LTRK 
selection in \LTRKLUM~\INVPB\ of Run II data, with \NWASOL\ events resulting 
in NWA solutions.
The KIN and PHI analyses are applied to the \DILEVT\ events of the DIL 
selection sample, corresponding to \DILLUM~\INVPB.
Of this sample, \KINSOL\ events pass kinematic reconstruction in the KIN
method, while the PHI analysis returns solutions for all \DILEVT\ events.
Each method applies the likelihood procedure described in 
Sec.~\ref{sec:procedure}, using the expected number of background events 
listed in Table~\ref{tab:peevents} for the LTRK or DIL selection sample 
after accounting for mass reconstruction efficiency.
As listed in Table~\ref{tab:results}, each likelihood fit returns
a constrained number of background events consistent with the expected
value.
The number of signal events returned from each likelihood fit ensures that
the total number of events in the likelihood agrees with that observed,
thereby accounting for the upward fluctuations in both the DIL and LTRK 
selection samples.


\begin{table}[b]
   \caption
   {
      Summary of results for the NWA, KIN, and PHI methods applied to the
      LTRK and DIL data samples.
      Listed for each method are:
      number of total observed events in the sample,
      number of events with mass solutions,
      expected number of background events,
      constrained likelihood fit values for signal and background events and
      top quark mass,
      and unconstrained likelihood mass.
   }
   \begin{ruledtabular}
   \begin{tabular}{cccccccc}

   {\bf method}
   & \multicolumn{3}{c}{\bf data sample}
   & \multicolumn{3}{c}{\bf constrained results}
   & {\bf unconstrained} \\

   & $N_{\rm tot}$
   & $N_{\rm sol}$
   & $n_{b}^{\rm exp}$
   & $n_{s}$
   & $n_{b}$
   & $m_t$ (\GEVCC)
   & $m_t$ (\GEVCC) \\ \hline

   {\bf NWA}
   & \LTRKEVT
   & \NWASOL
   & $14.1 \pm 3.5$
   & $32.4 \pm 7.4$
   & $13.4 \pm 3.5$
   & \NWAMASSSTAT
   & $168.3 \pm 4.9$ \\

   {\bf KIN}
   & \DILEVT
   & \KINSOL
   & $ 6.4 \pm 1.2$
   & $24.5 \pm 5.6$
   & $ 6.1 \pm 1.7$
   & \KINMASSSTAT
   & $168.4 \pm 6.1$ \\

   {\bf PHI}
   & \DILEVT
   & \PHISOL
   & $10.5 \pm 1.9$
   & $24.4 \pm 5.9$
   & $10.0 \pm 1.9$
   & \PHIMASSSTAT
   & $169.2 \pm 6.4$ \\

   \end{tabular}
   \end{ruledtabular}
   \label{tab:results}
\end{table}

\begin{figure}[tbp]
   \begin{center}
   \includegraphics[width={10cm}]{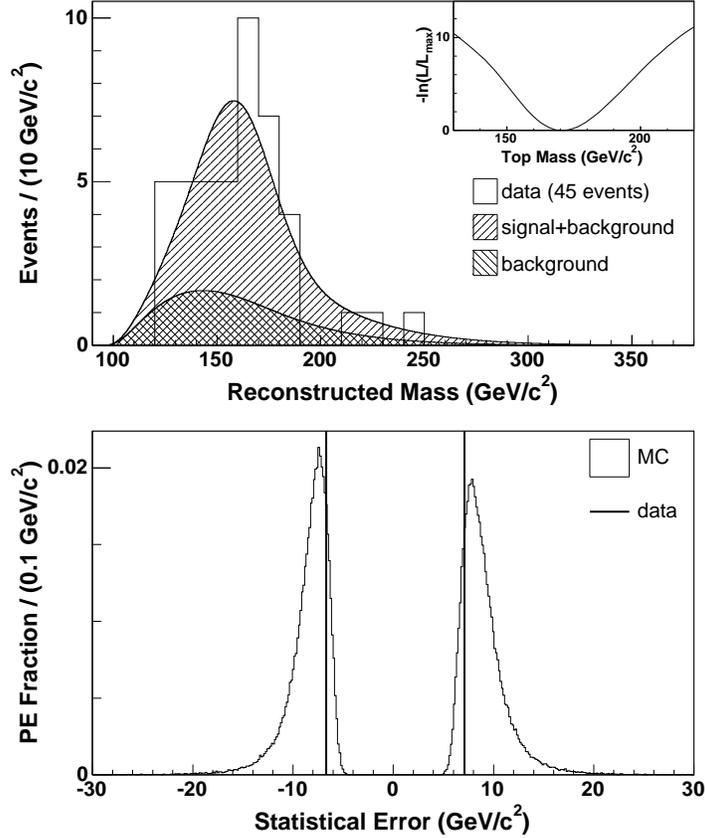}
   \caption
   {
      Results for the NWA method applied to the \LTRKEVT-event LTRK data
      sample, leading to \NWASOL\ solutions.
      Upper plot:
      reconstructed top quark mass for the data events (histogram),
      with normalized background and signal+background
      p.d.f.\ curves, and the likelihood function (inset).
      Lower plot:
      comparison of measured positive and negative statistical uncertainties
      in the data sample (vertical lines) with pseudo-experiments generated
      using the 170 \GEVCC\ signal template and assuming at least
      \NWASOL\ events observed.
   }
   \label{fig:nwa_results}
   \end{center}
\end{figure}
\begin{figure}[tbp]
   \begin{center}
   \includegraphics[width={10cm}]{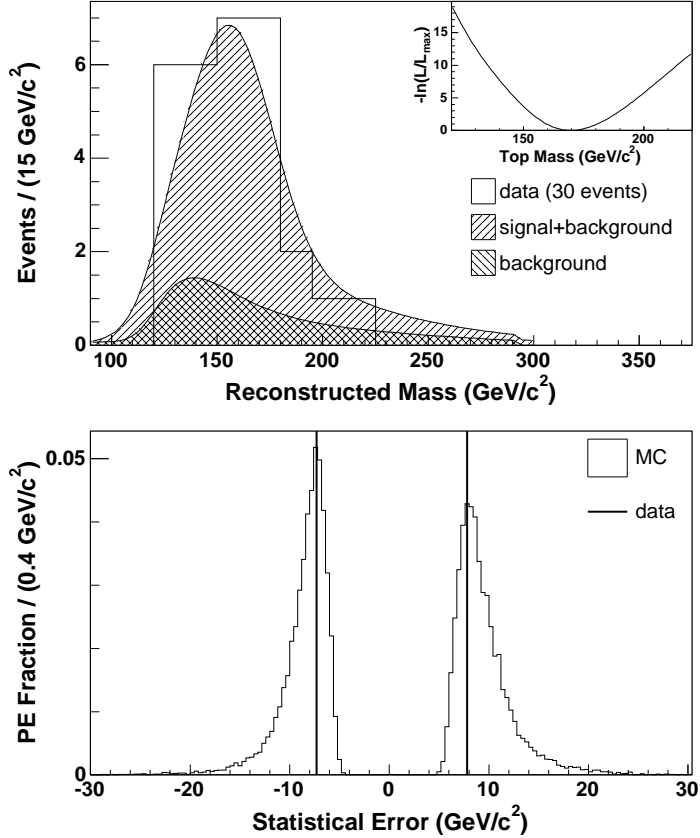}
   \caption
   {
      Results for the KIN method applied to the \DILEVT-event DIL data
      sample, leading to \KINSOL\ solutions.
      Upper plot:
      reconstructed top quark mass for the data events (histogram),
      with normalized background and signal+background
      p.d.f.\ curves, and the likelihood function (inset).
      Lower plot:
      comparison of measured average statistical uncertainty
      in the data sample (vertical line) with pseudo-experiments generated
      using the 170 \GEVCC\ signal template.
   }
   \label{fig:kin_results}
   \end{center}
\end{figure}
\begin{figure}[tbp]
   \begin{center}
   \includegraphics[width={10cm}]{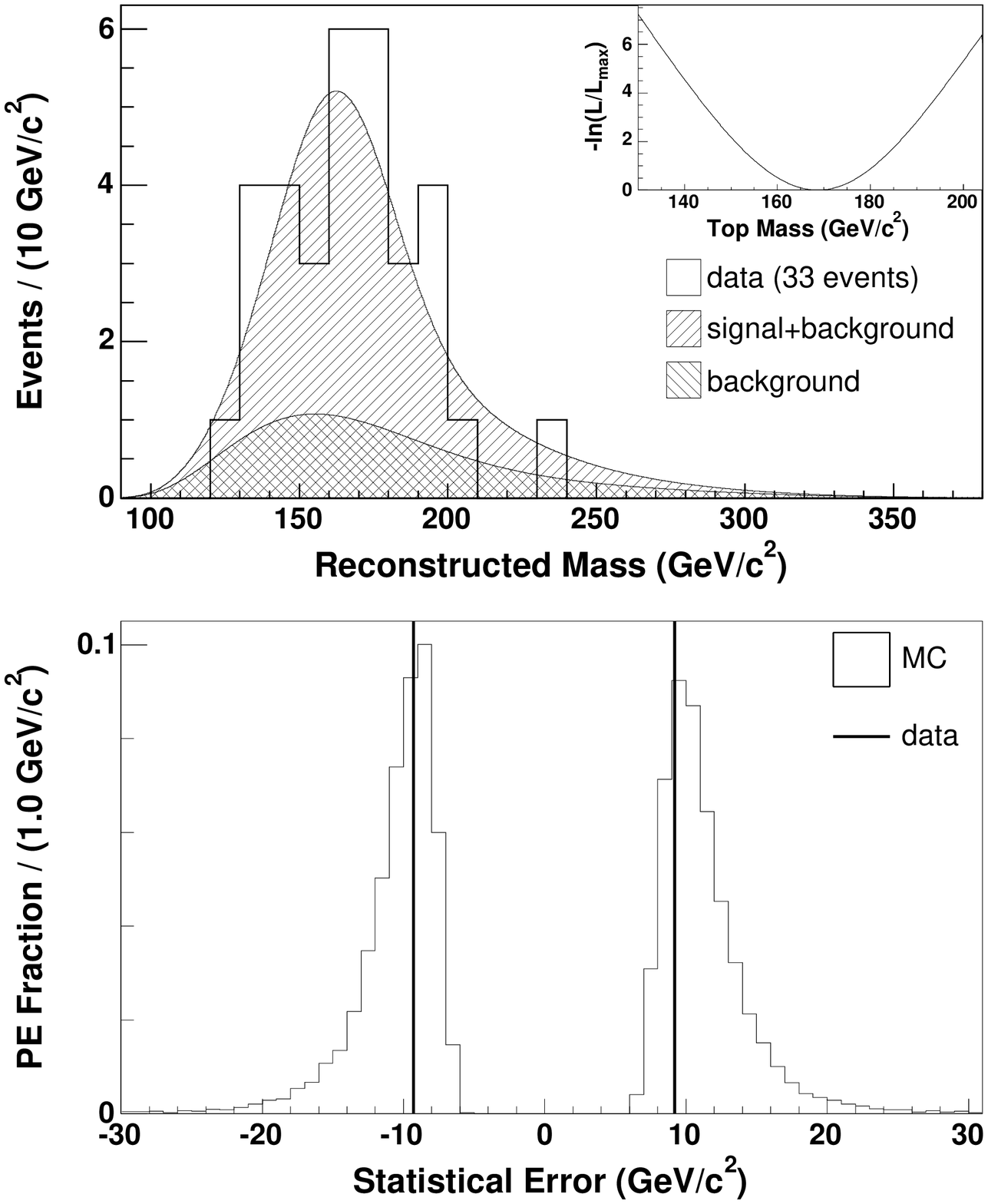}
   \caption
   {
      Results for the PHI method applied to the \DILEVT-event DIL data sample.
      Upper plot:
      reconstructed top quark mass for the data events (histogram),
      with normalized background and signal+background
      p.d.f.\ curves, and the likelihood function (inset).
      Lower plot:
      comparison of measured positive and negative statistical uncertainties
      in the data sample (vertical lines) with pseudo-experiments generated
      using the 170 \GEVCC\ signal template.
   }
   \label{fig:phi_results}
   \end{center}
\end{figure}

The upper plots of Figs.~\ref{fig:nwa_results}-\ref{fig:phi_results} 
show for the NWA, KIN, and PHI methods, respectively, the reconstructed 
top quark mass in the data, the normalized background and signal+background 
shapes, and the variation of
   $-\ln(\mathcal{L}/\mathcal{L}_{\rm max})$
as a function of the top quark mass hypothesis.
For each method, the final top quark mass is taken as the value 
of $m_t$ which minimizes the likelihood function.
Statistical uncertainties are obtained by taking the width at
   $-\ln(\mathcal{L}/\mathcal{L}_{\rm max}) + 0.5$,
and adjusting for the underestimation found in pull widths from
Figs.~\ref{fig:nwa_pulls}-\ref{fig:phi_pulls}.
Table~\ref{tab:results} summarizes the measured top quark mass and statistical
uncertainty for the three mass methods after pull width corrections.


The lower plots of Figs.~\ref{fig:nwa_results}-\ref{fig:phi_results} compare
the measured statistical uncertainties of the NWA, KIN, and PHI methods
with pseudo-experiments using the $m_t = 170$~\GEVCC\ sample which have the 
same number of events as that observed in the data for each method.
We find the probabilities for achieving the observed statistical uncertainties
to be
   9\%, 23\%, and 19\%
for the NWA, KIN, and PHI methods, respectively.
As a further cross-check, we remove the Gaussian constraint on the number
of background events in the likelihood procedure
({\it i.e.}, the term ${\mathcal L}_{n_b}$ in Eq.~\ref{eqn:ltotal}).
For all three methods, this unconstrained fit converges near zero background
events, which is found to occur for 21\% (NWA), 31\% (KIN), and 20\% (PHI) of
pseudo-experiments using the $m_t = 170$~\GEVCC\ sample.
The resulting top quark mass of the unconstrained fit, corrected for pull 
width, is consistent with the constrained result for each mass method (as 
seen in Table~\ref{tab:results}).
The statistical uncertainty returned by the unconstrained fit on the data 
sample is smaller than that of the constrained fit for all methods.
However, from studies of pseudo-experiments at $m_t = 170$~\GEVCC\ we expect
on average an improvement in statistical uncertainty of
   $1.1$, $1.5$, and $1.2$~\GEVCC\
for the NWA, KIN, and PHI methods, respectively, when applying the background
constraint to the likelihood function.


\clearpage
\section{\label{sec:systematics} Systematic Uncertainties }

Apart from the statistical uncertainty on the measured top quark mass due 
to the limited size of our data sample,
there are several sources of systematic uncertainty.
These systematic effects stem from uncertainty in the Monte Carlo simulation of
\TTBAR\ and background events, from mismodeling by the simulation
of the detector response to leptons and jets, and from the validity of various
assumptions made during the implementation of the mass measurement 
techniques.
As such, most sources of systematic uncertainty are common to all three mass 
analyses, and are estimated by adjusting a particular input value to the 
simulation and constructing new mass templates.
We then perform pseudo-experiments using events drawn from the new mass
templates, and compare the resulting median reconstructed top quark mass 
with that of the nominal simulation.
The sources of systematic uncertainty within each mass analysis are assumed 
to be uncorrelated, so that a total systematic uncertainty for each method is 
calculated as the sum in quadrature of the various sources,
as summarized in Table~\ref{tab:totalsyst}.

One of the largest sources of systematic uncertainty arises from potential 
mismodeling
of the jet energy measurement, through uncertainties in the various 
corrections applied to the measured jet energy~\cite{jetcor_NIM}.
These jet energy corrections involve the non-uniformity in response 
of the calorimeter as a function of $\eta$, effects from multiple \PPBAR\
collisions, the absolute jet energy scale for hadrons, energy deposition 
from the underlying \PPBAR\ event, and energy loss outside the jet search 
cone $\Delta R$.
A systematic uncertainty is estimated for each jet energy correction
by performing pseudo-experiments drawn from signal and background
templates with $\pm 1$ standard deviation in correction uncertainty,
and taking the half-difference in median reconstructed top quark mass between
the two results.
The uncertainties from each energy correction are then added in quadrature
to arive at a total systematic uncertainty on the jet energy scale.

Since the above jet energy corrections are developed from studies of
samples dominated by light-quark and gluon jets,
additional uncertainty occurs from extrapolating
this procedure to $b$-quarks.
The resulting systematic effect on jet energy is considered to stem from 
three main sources:
   uncertainty in the $b$-jet fragmentation model,
   differences in the energy response due to semi-leptonic decays
      of $b$-hadrons,
   and uncertainty in the color flow within top quark production and decay
      to $b$-jets~\cite{ljets_mass_PRD}.
As in the jet energy scale uncertainty, pseudo-experiments are performed 
on events where the $b$-jet energies have been altered by $\pm 1$ standard 
deviation for each uncertainty, and the resulting half-differences added
in quadrature to estimate the total systematic uncertainty due to $b$-jet 
energy uncertainty.

Several systematic uncertainties are due to the modeling of the \TTBAR\ 
signal.
We study the effects of the particular Monte Carlo generator chosen by 
comparing pseudo-experiments drawn from \PYTHIA\ simulation with events
taken from our nominal signal templates constructed using \HERWIG.
These generators differ in their hadronization models and in their handling
of the underlying \PPBAR\ event and multiple \PPBAR\ 
interactions~\cite{generators}.
We take the difference in reconstructed top quark mass 
between \HERWIG\ and \PYTHIA\ pseudo-experiments as the
systematic uncertainty due to choice of generator.
The systematic uncertainty associated with the initial state 
radiation (ISR) is studied by changing the QCD parameters for parton 
shower evolution according to comparisons between CDF Drell-Yan data and 
simulation~\cite{ljets_mass_PRD}.
Since final state radiation (FSR) shares the same Monte Carlo shower  
algorithms as ISR, these variations in QCD parameters are used to generate
FSR systematic samples by varying a set of parameters specific to FSR
modeling.
We then compare the reconstructed top quark mass from samples with 
increased and decreased ISR and FSR to estimate the systematic uncertainty 
due to these sources.
The uncertainty in reconstructed top quark mass from our choice of parton
distribution function (PDF) is found by comparing two different groups
(\texttt{CTEQ5L}~\cite{CTEQ} and \texttt{MRST72}~\cite{MRST}).
Additionally, \texttt{MRST72} and \texttt{MRST75} sets, derived using 
different $\Lambda_{QCD}$ values, are compared, and 20 eigenvectors 
within the \texttt{CTEQ6M} group are 
varied by $\pm 1$ standard deviation.
Differences in pseudo-experiment results from these variations are added
in quadrature to arive at a total systematic uncertainty from the choice
of PDF.
Further studies comparing LO with NLO \TTBAR\ Monte Carlo show a negligible
effect on the reconstructed top quark mass.

Since our background template is also derived from simulation, another
source of systematic uncertainty reflects the potential mismodeling by
the Monte Carlo of the background shape.
Background events may pass event selection through processes which are not
accurately modeled in the simulation, such as tracks or jets passing through
gaps in detector elements.
This uncertainty is estimated by measuring the resulting top quark mass from
pseudo-experiments where the track $p_T$, jet energy, and \MET\ of each 
background sample have been altered by the measured discrepancies in these
quantities between simulation and data.
Another systematic uncertainty affecting the background shape is due to
uncertainty in the relative composition of the background sources used
to construct the total background template.
To estimate this uncertainty, we measure the effect on top quark mass from
pseudo-experiments where the relative combination of Drell-Yan and fake
backgrounds (the largest two sources) is adjusted by predicted 
uncertainty.

The finite statistics in the simulated signal and background templates 
result in a systematic uncertainty on the parameterized p.d.f's used in 
the likelihood (Eq.~\ref{eqn:lshape}), even if modeling of the 
signal and background processes is correct.
As described in Section~\ref{sec:procedure},
the PHI method accounts for this uncertainty in template parameterizations
within the statistical uncertainty returned by the likelihood minimization
through the term
   $\mathcal{L}_{\rm param}$
of Eq.~\ref{eqn:lparam}.
The NWA and KIN procedures estimate directly the top quark mass uncertainty
due to finite template statistics, and incorporate this effect into the
total systematic uncertainty.
For each signal template, we Poisson fluctuate the number of events in each
bin to create a new template, which is parameterized according to
Eqs.~\ref{eqn:paramsig} and~\ref{eqn:paramkin}.
We then perform pseudo-experiments drawing signal events from the nominal
templates but applying them to a likelihood fit with the fluctuated signal
p.d.f.\ in Eq.~\ref{eqn:lshape}, producing a distribution of reconstructed
top quark mass.
Repeating this procedure many times, we estimate the systematic uncertainty
due to limited statistics in the signal templates as the root mean square 
of the median reconstructed top quark masses from the fluctuated 
pseudo-experiments.
In a similar fashion, we estimate the analogous systematic uncertainty
due to limited
background template statistics by fluctuating each template bin of the 
various background components.


\begin{table}
   \caption
   {
      Summary of the systematic uncertainties on the top quark mass
      measurement (in \GEVCC) for the NWA, KIN, and PHI analyses.
      The total uncertainty is obtained by adding the individual contributions
      in quadrature.
      (The uncertainty due to signal and background template statistics
      for the PHI method is accounted for in the total statistical 
      uncertainty).
   }
   \begin{ruledtabular}
   \begin{tabular}{lccc}

   {\bf systematic source} & {\bf NWA} & {\bf KIN} & {\bf PHI} \\ \hline

   Jet energy scale          &  3.4       &  3.2       &  3.5  \\
   $b$-jet energy            &  0.6       &  0.6       &  0.7  \\
   MC generator              &  0.5       &  0.6       &  0.7  \\
   PDF's                     &  0.5       &  0.5       &  0.6  \\
   ISR                       &  0.6       &  0.6       &  0.6  \\
   FSR                       &  0.5       &  0.3       &  0.4  \\
   Background shape          &  2.6       &  1.6       &  1.5  \\
   \hline
   Template statistics       &            &            &       \\
   \hspace*{10pt} Signal     &  0.2       &  0.4       &  n/a  \\
   \hspace*{10pt} Background &  1.3       &  1.2       &  n/a  \\
   \hline
   Total                     &  \NWASYST  &  \KINSYST  &  \PHISYST  \\ \hline

   \end{tabular}
   \end{ruledtabular}
   \label{tab:totalsyst}
\end{table}


\clearpage
\section{\label{sec:combo} Combination of Measurements }

Table~\ref{tab:combo} shows the results, including statistical and
systematic uncertainties, for the NWA, KIN, and PHI analyses.
The three results are consistent, and can be combined to improve
upon the overall precision of the top quark mass measurements using the
template method.
The combination procedure follows the Best Linear Unbiased Estimation (BLUE) 
method~\cite{BLUE}.
In this technique, the final result consists of a linear combination of the
individual measurements.
The measured statistical and systematic uncertainties for
each measurement, along with their correlations, are used to
construct an error matrix, which upon inversion gives the corresponding
weights for each method within the combined result.

The statistical correlations between methods are determined from
simulated samples over a range of top quark masses from 
   155 to 195~\GEVCC.
Pseudo-experiments from these samples, each corresponding to an integrated 
luminosity of
   $350$~\INVPB,
are constructed.
The LTRK and DIL selection criteria are applied to the pseudo-experiments
to model the expected signal sample, as well as correlations between the
selection methods.
Simulated background events are added to each pseudo-experiment according to
the expected contributions to each selection from the three background sources
(as listed in Table~\ref{tab:selections}).
Based on studies from simulation, the 
  \WJETS\ fake background 
is assumed to be uncorrelated between the DIL and LTRK selections.
For the diboson and Drell-Yan DIL backgrounds, a Poisson fluctuated number of
events with mean
   $\langle N_{\rm DIL} \rangle$
is added to each pseudo-experiment, drawn from a pool of events passing the
DIL selection.
The LTRK diboson and Drell-Yan backgrounds are constructed by taking into 
account the expected number of common events,
   $N_{\rm LTRK \cdot DIL}$,
between the two selections, as determined from simulation.
These LTRK backgrounds are thus the union of
   $N_{\rm LTRK \cdot DIL}$
common events with a number of events
   $\langle N_{\rm LTRK}-N_{\rm LTRK \cdot DIL} \rangle$
drawn from the pool of those events passing only the LTRK selection.
From these pseudo-experiments of signal and background events, we find an 
overlap between the LTRK and DIL selections of approximately
   $30\%$,
compared with the \COMOVERLAP\ overlap observed in the two data samples.

The three measurement methods are applied to each pseudo-experiment,
returning a reconstructed top quark mass and expected statistical
uncertainty.
Pseudo-experiments where the likelihood minimization for any of the 
methods fails to converge are removed from consideration.
This situation occurs for approximately 0.8\% of all pseudo-experiments.
Also, pseudo-experiments must have a returned mass which lies within 
the likelihood minimization mass range of all methods.
The correlation coefficient between two mass measurements $m_i$ and $m_j$
is calculated from the $N$ pseudo-experiments at each mass sample by the 
equation:
\begin{eqnarray}
   \rho_{ij} = \frac
   { N \sum m_i m_j - \sum m_i \sum m_j }
   { \sqrt { N \sum m_i^2 - ( \sum m_i )^2 }
     \sqrt { N \sum m_j^2 - ( \sum m_j )^2 } }
   .
\end{eqnarray}
These statistical correlations are shown in Table~\ref{tab:combo} for the
   $m_t = 170$~\GEVCC\
sample, and are stable across the 155 to 195~\GEVCC\ mass samples.
Systematic uncertainties common to all three methods, as listed in 
Table~\ref{tab:totalsyst}, are assumed to be 100\% correlated, while the
systematic uncertainties due to limited template statistics for the NWA and 
KIN methods are assumed to be uncorrelated.
After incorporating systematic effects, Table~\ref{tab:combo} shows the 
total correlations between methods in parentheses, and the resulting weights
of the methods after inversion of the constructed error matrix.

Since measurements producing smaller mass values tend to have correspondingly
smaller statistical uncertainties, an iterative combination procedure is
performed in order to prevent bias towards lower mass values.
In this procedure, each measurement method is assumed to have a constant
fractional statistical uncertainty (taken from the expected uncertainty at
   $m_t = 178$~\GEVCC\
in Table~\ref{tab:peevents}).
Combining the three measurements, each method's statistical uncertainty is
then extrapolated to the combination value, and the procedure repeated until
the combined result converges.

\begin{figure}[tbp]
   \begin{center}
   \includegraphics[width={10cm}]{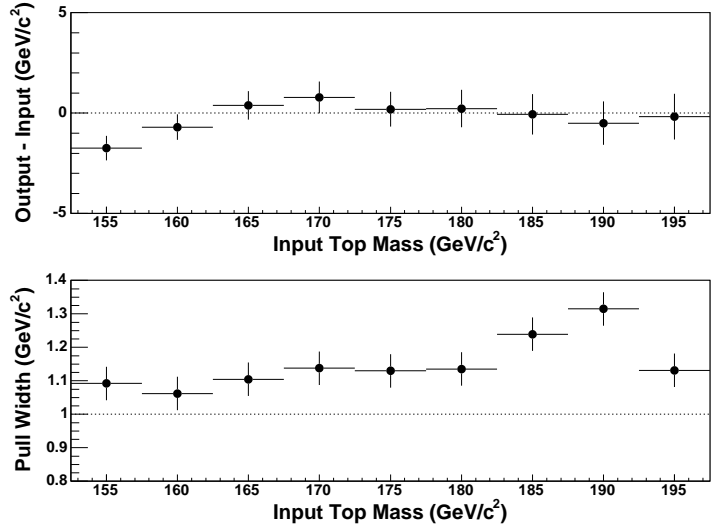}
   \caption
   {
      Summary of the difference between output and input top quark mass
      (upper) and width of pull distributions (lower) 
      for the combination pseudo-experiments, 
      as a function of generated input top quark mass.
   }
   \label{fig:com_pulls} 
   \end{center}
\end{figure}

Potential bias in the combination technique is studied by using the 
pseudo-experiment results from each method as input.
As shown in Fig.~\ref{fig:com_pulls}, we observe no bias in the 
residual difference between the input and output top quark mass
of the combination result for samples above
   $m_t = 160$~\GEVCC.
However, non-unity pull widths for all mass samples indicate that the 
statistical uncertainty of the combination procedure is slightly 
underestimated.
Studies of toy Monte Carlo pseudo-experiments suggest that this
underestimation may
stem from deviations in the data sample from assumptions made in the 
BLUE method ({\it e.g.}, Gaussian uncertainties).
Applying the combination procedure to the NWA, KIN, and PHI data sample
measurements, and correcting the returned statistical uncertainty by the 
average pull width over all mass samples (a scale factor of 1.15),
yields a top quark mass of
   \COMMASSSYST~\GEVCC.
The close proximity of the three template measurements with respect
to their measured uncertainties leads to a $\chi^2$ per degrees of freedom
for the combination of $\chi^2/{\rm d.o.f.} = 0.017/2$, 
corresponding to a $p$-value of 99\%.

\begin{table}[tbp]
   \caption
   {
      Summary of measured top quark masses for the
         Neutrino Weighting Algorithm (NWA),
         Full Kinematic Analysis (KIN),
         and Neutrino $\phi$ Weighting Method (PHI),
      along with their statistical (total) correlations, and contributing 
      weight to the combined top quark mass result.
   }
   \begin{ruledtabular}
   \begin{tabular}{cccccc}

   {\bf method}
      & {\bf result ($\bf GeV/c^2$)}
      & \multicolumn{3}{c}{\bf correlation}
      & {\bf weight} \\

   {\bf }
      &
      & {\bf NWA}
      & {\bf KIN}
      & {\bf PHI}
      &      \\ \hline

   {\bf NWA} & \NWAMASSSYST & 1.00 (1.00) & 0.14 (0.32) & 0.25 (0.40) & 47\% \\
   {\bf KIN} & \KINMASSSYST &             & 1.00 (1.00) & 0.35 (0.46) & 38\% \\
   {\bf PHI} & \PHIMASSSYST &             &             & 1.00 (1.00) & 15\% \\

   \end{tabular}
   \end{ruledtabular}
   \label{tab:combo}
\end{table}


\clearpage
\section{\label{sec:summary} Summary }

We have performed three separate measurements of the top quark mass
from \TTBAR\ events produced in \PPBAR\ collisions at a center-of-mass 
energy of 1.96 TeV, using the Run II Collider Detector at Fermilab.
The mass measurements employ one of two complementary selection algorithms 
to extract \TTBAR\ events
where both $W$ bosons from the top quarks decay into leptons
($e\nu, \mu \nu$, or $\tau \nu$), producing data samples of \DILLUM\ and
\LTRKLUM~\INVPB.
Each measurement technique determines a single top quark mass for an 
event by making different assumptions in order to resolve the underconstrained 
dilepton \TTBAR\ decays.
For each method, template mass distributions are constructed from simulated
signal and background processes, and parameterized to form continuous
probability density functions.
A likelihood fit incorporating these parameterized templates is then performed
on the data sample masses in order to derive a final top quark mass.

One method, the Neutrino Weighting Algorithm (NWA), measures a top quark 
mass of 
   \NWAMASSSYST~\GEVCC.
A second technique, called the Full Kinematic Analysis (KIN), results in
a mass measurement of
   \KINMASSSYST~\GEVCC.
A third analysis using the Neutrino $\phi$ Weighting Method (PHI) measures
a value of 
   \PHIMASSSYST~\GEVCC.
Accounting for correlations in the statistical and systematic uncertainties
between methods, we combine the three results, giving a top quark mass in the
dilepton channel of
   \COMMASSSYST~\GEVCC.
This combined result is consistent with the CDF Run II ``lepton+jets'' 
channel~\cite{ljets_mass_PRL}, which used a
   318~\INVPB\
data sample to measure a top quark mass of
   $173.5^{+3.9}_{-3.8}$~\GEVCC,
and thus gives no indication of new physics in the dilepton channel.
The three template analyses are also consistent with 
a fourth CDF top quark mass measurement in the dilepton 
channel~\cite{dilepton_mass_PRL}, which applies a matrix-element technique
to the \DILEVT-event DIL selection sample.



\begin{acknowledgments}
   We thank the Fermilab staff and the technical staffs of the participating 
institutions for their vital contributions. This work was supported by 
the U.S. Department of Energy and National Science Foundation; the 
Italian Istituto Nazionale di Fisica Nucleare; the Ministry of Education, 
Culture, Sports, Science and Technology of Japan; the Natural Sciences 
and Engineering Research Council of Canada; the National Science Council 
of the Republic of China; the Swiss National Science Foundation; the A.P. 
Sloan Foundation; the Bundesministerium f\"ur Bildung und Forschung, 
Germany; the Korean Science and Engineering Foundation and the Korean 
Research Foundation; the Particle Physics and Astronomy Research Council 
and the Royal Society, UK; the Russian Foundation for Basic Research; 
the Comisi\'on Interministerial de Ciencia y Tecnolog\'{\i}a, Spain; in 
part by the European Community's Human Potential Programme under contract 
HPRN-CT-2002-00292; and the Academy of Finland. 

\end{acknowledgments}



\end{document}